\documentclass[pra,nofootinbib,floats,aps,twocolumn,tightenlines,superscriptaddress]{revtex4-1}

\usepackage{graphicx}
\usepackage{bbold}
\usepackage{mathtools}

\usepackage{hyperref}

\usepackage{bm}

\usepackage{amsfonts,amsmath,amssymb,amsthm}

\DeclareMathOperator{\Tr}{Tr}

\allowdisplaybreaks

\newcommand{\ii}{\mathrm{i}}

\newcommand{\ket}[1]{\left| {#1} \right\rangle}
\newcommand{\bra}[1]{\left\langle {#1} \right|}

\newcommand{\braket}[2]{\left\langle {#1}\left|{#2}\right.\right\rangle}
\newcommand{\proj}[2]{\left| {#1} \right\rangle\!\left\langle {#2} \right|}
\renewcommand{\Re}[1]{\operatorname{Re}{#1}}

\newcommand{\tr}{\text{Tr}}
\newcommand{\norm}[1]{\left|\left|#1\right|\right|}
\renewcommand{\d}[0]{\mathrm{d}}

\usepackage[usenames,dvipsnames]{xcolor}

\begin{abstract}
Randomness is an indispensable resource in modern science and information technology. Fortunately, an experimentally simple procedure exists to generate randomness with well-characterized devices: measuring a quantum system in a basis complementary to its preparation. Towards realizing this goal one may consider using atoms or superconducting qubits, promising candidates for quantum information processing. However, their unavoidable interaction with the electromagnetic field affects their dynamics. At large time scales, this can result in decoherence. Smaller time scales in principle avoid this problem, but may not be well analysed under the usual rotating wave and single-mode approximation (RWA and SMA) which break the relativistic nature of quantum field theory. Here, we use a fully relativistic analysis to quantify the information that an adversary with access to the field could get on the result of an atomic measurement. % information that an atom in contact with a field shares with it at short time scales. We quantify this information in terms of the guessing probability of an adversary having access to the field. 
Surprisingly, we find that the adversary's guessing probability is not minimized for atoms initially prepared in the ground state (an intuition derived from the RWA and SMA model).
\end{abstract}

\begin{document}
\title{Certified Randomness from a Two-Level System in a Relativistic Quantum Field}
%\title{The impact of relativity on randomness generation in quantum optics}
%\title{Randomness generation from atomic detectors in quantum fields}

\author{Le Phuc Thinh}
\affiliation{Centre for Quantum Technologies, National University of Singapore, 3 Science Drive 2, Singapore 117543}
\affiliation{QuTech, Delft University of Technology, Lorentzweg 1, 2628 CJ Delft, The Netherlands}

\author{Jean-Daniel Bancal}
\affiliation{Centre for Quantum Technologies, National University of Singapore, 3 Science Drive 2, Singapore 117543}

\author{Eduardo Mart\'in-Mart\'inez}
\affiliation{Institute for Quantum Computing, University of Waterloo, Waterloo, Ontario, N2L 3G1, Canada}
\affiliation{Department of Applied Mathematics, University of Waterloo, Waterloo, Ontario, N2L 3G1, Canada}
\affiliation{Perimeter Institute for Theoretical Physics, Waterloo, Ontario N2L 2Y5, Canada}

\maketitle

\section{Introduction\label{sec:intro}}
Randomness is a fundamental resource for tasks as varied as numerical simulations, cryptography, algorithms or gambling~\cite{Metropolis1949,Motwani1995}. It is known that quantum systems can be used to generate truly unpredictable outcomes. While measurements on entangled states allow one to certify this randomness under a small set of assumptions~\cite{Pironio2010}, measurements on single systems can already produce certified randomness if a higher ``level of characterization" is taken into consideration~\cite{Law14}. Here, we consider the randomness that can be certified by measuring a single atom in the latter case.

Atoms do not exist isolated: They always, and unavoidably, interact with the electromagnetic field. If we want to use an atomic system as a source of randomness, for example by preparing a state in one basis and then measuring in a mutually unbiased basis, one has to consider that between the time of preparation ($t=0$) and the time of measurement ($t=T$), the atom interacts with the field, thus effectively sharing some information with the field. If this information can be retrieved by an adversary having access to the field at a later time, it may compromise the unpredictability of the atom's measurement result.

When the time between preparation and measurement is large decoherence may leave the atom in a mixed state, thus significantly impacting the efficiency of an atomic random number generator. One could hope to circumvent this problem by considering a short time between preparation and measurement. However, certifying randomness in this regime requires special care since relativistic effects %(which come about through non-rotating wave, non-single-mode contributions \cite{EMM2015}) 
are expected to influence the leading order contributions to the correlations between the atom and the field in this situation, in a similar manner as in the case of entanglement harvesting \cite{Valentini1991,Rez03,RRS05,Pozas2015}.

It has been discussed in the context of relativistic quantum information that atomic probes which interact with the electromagnetic field become, in general, entangled with these fields. This is true even when the dynamics of the atom-field system is dominated by vacuum fluctuations \cite{RRS05,Pozas2015}. These correlations are neglected in quantum optics when working under the usual rotating wave approximation (RWA) and the single mode approximation (SMA) \cite{ScullyBook} -- two approximations which break the Lorentz covariance of the interaction theory and allow for causality violations and superluminal signalling~\cite{EMM2015}. However, since such correlations could be used by an adversary to guess the result of the atomic measurement, neglecting them potentially results in an underestimate of the adversary's power.

In this article, we focus on the regime of short time between preparation and measurement, and take into account the fully relativistic\footnote{By relativistic, we mean here that the detector is locally coupled to a Lorentz covariant field. This excludes any possibility of superluminal signalling (present within the SMA and RWA)~\cite{EMM2015} and guarantees a proper description of high frequency modes relevant at short times.} light-matter interaction model. Our analysis applies for instance to the case of an atomic probe in an optical cavity or free space, or to a superconducting qubit coupled to a transmission line.

We show that, even for atoms in the ground state in the presence of vacuum, the field fluctuations drive the creation of field-atom entanglement at a significant level. This implies, perhaps contrary to intuition, that reducing the time from preparation to measurement generally does not spare a decrease in the randomness extractable from the atom, even for extremely short timescales. We hence conclude that relativistic effects need to be taken into account in the short time regime. 

We also show that, even for relatively long waiting times between preparation and measurement, the ground state of the atom together with the vacuum state of the field is not the optimal state for randomness extraction when all relativistic considerations are factored in. This contradicts the intuition stemming from the SMA and RWA according to which, if an atom starts in its ground state and the field is not excited, then the atom would not get entangled with the field, and so it would share no information with the field. Thus, our results demonstrate that the actual behavior is really different from the one given by these usual approximations. Quantitatively, for typical timescales and coupling regimes of strong and ultra-strong coupling in quantum optics and superconducting qubits in transmission lines, we estimate that that the use of the SMA and RWA leads to an overestimation of the amount of randomness that can reach magnitudes of the order of 10\%.

\section{Quantifying the randomness extractable from an atomic detector\label{sec:main}}

\begin{figure}
\centering
\includegraphics[width=0.45\textwidth]{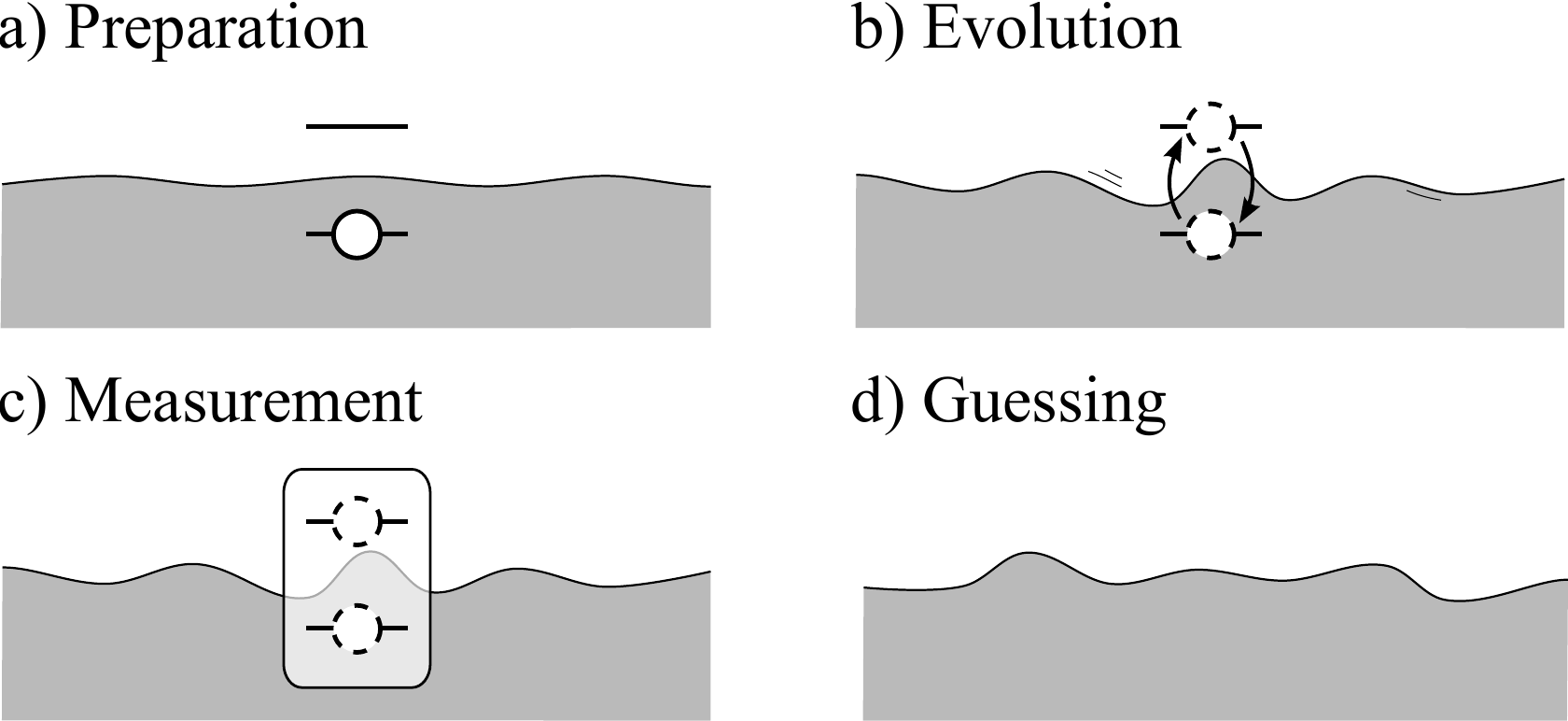}
\caption{Certifying randomness of an atomic measurement after interacting with a field. a) Preparation: A two-level atom is prepared in some state (here the ground state) and starts interacting with in an empty field. b) Evolution: the atom and the field evolve for a time $T$. c) Atomic measurement: a random outcome is obtained by measuring the state of the atom in an appropriate basis. d) Guessing: the adversary can access the field (possibly at a later time) to try to guess the result of the atomic measurement.}
\label{fig:scheme}
\end{figure}

We consider the situation in which a user wants to generate random bits by performing a quantum measurement on an atom. For this purpose, he prepares the atom in a state $\ket{\psi_A}$, and then performs an \textit{optimal} von Neumann measurement on it (e.g. in a complementary basis)\footnote{We leave the question of performing more general POVMs, possibly by involving additional ancillas~\cite{Law14}, for further study. This could potentially certify up to two bits of randomness per measurement~\cite{Acin15}.}. Since the measurement is not performed simultaneously with the state preparation, this leaves some time $T$ for the atom to interact with the electromagnetic field between its preparation and measurement (c.f. Fig.~\ref{fig:scheme}). In particular, this interaction modifies the optimal measurement to be performed at time $T$ with respect to the initial mutually unbiased measurement.

Typically, this joint evolution results in the state of the atom and the field being partially entangled. After this interaction, the field thus contains some information about the outcome observed by the user upon measurement of the atom. Assuming that the field is not fully under control of the user, but can eventually be accessed by someone interested in guessing the outcome of the atom measurement (i.e. an adversary), one must evaluate how much information about the atom's state was shared with the field during this interaction time $T$ in order to certify the amount of randomness that can be extracted from the atom's measurement. We now describe this computation.\ \\

Let us consider a two-level atom and a massless scalar field ${\phi}(x,t)$ in 1+1 dimensions initially prepared in the state $\rho_i=\proj{\psi_A}{\psi_A}\otimes\proj{0}{0}$. We model the atom-field interaction via a derivative coupling given by the following interaction Hamiltonian in the interaction picture
\begin{equation}
\label{eq:int}
H_I(t) = \lambda \int \d x\, F(x-x_a)\chi(t){\mu}(t)\partial_t{\phi}(x,t).
\end{equation}
where $\lambda$ is the coupling strength, $F(x-x_a)$ the spatial profile of atom positioned at $x_a$ (henceforth assumed symmetric about $x_a$), $\chi(t)$ the coupling switching function and ${\mu}(t)=(e^{\ii \Omega t }  \sigma^++e^{-\ii \Omega t }  \sigma^-)$ the atom's monopole moment. This is a simplified version of the light-matter interaction --- it can be thought of as a polarization-insensitive direct coupling to the electric field which is the derivative of the vector potential $\bm E=\partial_t\bm A$ in a 1D cavity such an optical fibre. The derivative coupling has been employed in the past to ameliorate the IR behaviour of the model in many different contexts \cite{IRdet,Wang,Benitoooo}. In our case, the use of this model also allows us to minimize the impact of neglecting the zero-mode dynamics in case of the periodic cavity \cite{zeromode}. While simple, this family of Unruh-DeWitt detector models have been proved to capture the fundamental features of the light-matter interactions \cite{Alvaro,Wavepackets}. 

Notice, however, that despite its simplicity, the model we consider here fully describes the phenomenology of the light-matter interaction \cite{ScullyBook} assuming neither the RWA nor the SMA. As a consequence, this interaction model is a causally well-behaved theory \cite{EMM2015}. This is crucial in the current context, where we expect vacuum correlations to play a role in the amount of randomness that can be extracted by measuring an atomic system in short times after preparation. Also, notice that the model does not consider the atom as a point-like particle but incorporates its spacial profile. Although the results are largely independent of the particular profile of the detector, its inclusion makes the analysis more general.

%Note also that a model making only one of the SMA and RWA may incorporate unphysical features (see discussion in Sec.~\ref{}).

After the interaction with the field, the global state is given by
\begin{align*}
\rho_{AF} = \proj{\psi_{AF}}{\psi_{AF}}=U\rho_iU^{\dagger}, \\
U = \mathcal{T}\exp\left(-\ii\int_{-\infty}^{\infty}\d t H_I(t)\right)
\end{align*}
where $\mathcal{T}$ represents time ordering. 

After a time $T$, the atom is measured in some basis. The global pure state of the atom and field then effectively `collapses' into a state of the form $\rho_{XF}^x=\proj{x}{x}\otimes\tau^x_F$ with probability $p_X(x)$. Here, $x$ is the result of the measurement and 
$\tau^x_F=\Tr_{A}(P_x\ket{\psi}_{AF}\!\bra{\psi})/\Tr{(P_x\ket{\psi}_{AF}\!\bra{\psi}})$ is the state in which the \textit{field} is left when the measurement result is $x$, for a von Neumann measurement $\{P_x\}$. This part of the state is the one that an adversary Eve could get in contact with, and eventually measure in order to infer the value of $x$.

The amount of randomness that can be extracted from the outcome of the measurement performed at time $T$ with respect to an adversary having access to the quantum field can be quantified by the conditional min-entropy
\begin{equation}
H_{\min}(X|F)_{\rho_{XF}} = -\log P_g(X|F)_{\rho_{XF}},
\end{equation}
where $P_g(X|F)_{\rho_{XF}}$ is the probability that the outcome  (random variable) $X$ is guessed correctly given the state of the quantum field $F$, and $\rho_{XF}=\sum_x p_X(x)\rho^x_{XF}$. Note that from a mathematical standpoint, the infinite-dimensionality of the quantum field as side information may a priori require some special care~\cite{BFS11}. The interpretation of the min-entropy in this context, as well as its characterizing properties, remain however intact. Using the invariance of the conditional min-entropy under local isometries and the fact that the atom under consideration can only be excited in a finite number of levels, we can effectively treat the quantum field $F$ as a finite dimensional system. For this, we consider the state $\ket{\psi_{AF}}$ of the atom and field just before the von Neumann measurement. By the Schmidt decomposition, there exists a basis of the quantum field $\{\ket{f_0},\ket{f_1}\}$ in which this state can be written as $\ket{\psi}_{AF}=\sqrt{\lambda_0}\ket{0f_0}+\sqrt{\lambda_1}\ket{1f_1}$. An isometry can be set up between the field $F$ and an arbitrary qubit $E$ of Eve so that all the entanglement between $A$ and $F$ can be transferred to $\ket{\psi}_{AE}=\sqrt{\lambda_0}\ket{00}+\sqrt{\lambda_1}\ket{11}$. We can thus compute the min-entropy on $\rho_{XE}=\sum_x p_X(x)\proj{x}{x}\otimes\tau^x_E$ where $\tau^x_E=\Tr_{A}{(P_x\ket{\psi}_{AE}\bra{\psi}})/\Tr{(P_x\ket{\psi}_{AE}\bra{\psi}})$ is the \textit{qubit} state hold by Eve whenever the atom is projected into outcome $x$.

To arrive at an analytic expression for the min-entropy, we recall two facts. First, the guessing probability $P_g$ for cq states can be interpreted as the optimal success probability for Eve to distinguish the (normalized) ensemble of states $\{\tau^x_E\}$:
\begin{align*}
P_g(X|E)_{\rho_{XE}} &= \max_{\mathcal{E}} \sum_x p_X(x)\bra{x}\mathcal{E}(\tau^x_E)\ket{x} \\*
&= \max_{\Pi_x} \sum_x p_X(x)\tr(\Pi_x\tau^x_E),
\end{align*}
where optimizing over TPCPMs $\mathcal{E}$ is equivalent to optimizing over POVMs $\{\Pi_x=\mathcal{E}^\dagger(\proj{x}{x})\}$. Second, the optimal success probability for distinguishing an ensemble consisting of only two states is given be the Holevo-Helstrom theorem. Hence we find that the conditional min-entropy is given by
\begin{equation*}
H_{\min}(X|E) = -\log\left[\frac{1}{2} + \frac{1}{2}\norm{p_X(0)\tau^0_E-p_X(1)\tau^1_E}_1\right].
\end{equation*}

The measurement providing the largest amount of randomness from the atom can be found by optimization over all von Neumann measurements, namely $H_{\min}^*(X|E)=\max_{\{P_x\}} H_{\min}(X|E)_{\rho_{XE}}$. One can check that the result of this optimization can be expressed in terms of the purity $\Tr(\rho_A^2)$
%eigenvalues $\lambda_0,\lambda_1$
of the reduced density-matrix $\rho_A$ only as
\begin{equation}
\label{eq:randopt}
%H_{\min}^*(X|E)= -\log\left[\frac{1}{2} + \frac{1}{2}\sqrt{1-(\lambda_0-\lambda_1)^2}\right].
H_{\min}^*(X|E)= -\log\left[\frac{1}{2} + \sqrt{\frac{1-\Tr(\rho_A^2)}{2}}\right].
\end{equation}
To see this, note that given the assumed form of $\ket{\psi}_{AE}$ and orthogonal projection $P_0:=\proj{m_0}{m_0}, P_1:=\proj{m_1}{m_1}$ with $\ket{m_0}=\cos\theta\ket{0}+e^{\ii \phi}\sin\theta\ket{1}$, $\ket{m_1}=\sin\theta\ket{0}-e^{\ii \phi}\cos\theta\ket{1}$, the operators $p_X(0)\tau^0_E=\proj{e_0}{e_0}$ and $p_X(0)\tau^1_E=\proj{e_1}{e_1}$ can be explicitly computed
\begin{align*}
\ket{e_0}=\sqrt{\lambda_0}\braket{m_0}{0}\ket{0}+\sqrt{\lambda_1}\braket{m_0}{1}\ket{1}\\
\ket{e_1}=\sqrt{\lambda_0}\braket{m_1}{0}\ket{0}+\sqrt{\lambda_1}\braket{m_1}{1}\ket{1}
\end{align*}
which gives
\begin{equation*}
\norm{p_X(0)\tau^0_E-p_X(1)\tau^1_E}_1=\sqrt{1-4|\braket{e_0}{e_1}|^2}.
\end{equation*}
Finally, it is useful to note that the fidelity between $\ket{e_0}$ and $\ket{e_1}$ reaches its maximum at $(\lambda_0-\lambda_1)^2/4=\Tr(\rho_A^2)/2-1/4$.

The computation of the conditional min-entropy thus reduces to a computation of the reduced atomic state after the interaction with the quantum field. This is the subject of the next subsections.

\subsection{The final atomic state from perturbation theory}

For small enough values of the coupling strength $\lambda$, the time-evolved density matrix is well approximated by the following perturbative expansion:
\begin{equation}\label{pertexp}
\rho \simeq \rho_i + \rho^{(1)} + \rho^{(2)},
\end{equation}
where $\rho^{(1)}=U^{(1)}\rho_i + \rho_iU^{(1)\dagger}$ and $\rho^{(2)}=U^{(1)}\rho_iU^{(1)\dagger} + U^{(2)}\rho_i + \rho_iU^{(2)\dagger}$ are the first and second order perturbation terms in $\lambda$, and
\begin{align}\label{afterfour}
\notag U^{(1)} &= -\ii\int_{-\infty}^{\infty}\!\!\!\d t\, H_I(t), \\
U^{(2)} &= -\int_{-\infty}^{\infty}\!\!\!\d t\int_{-\infty}^{t}\!\!\d t' H_I(t)H_I(t').
\end{align}
Since we are going to consider three different boundary condition scenarios (free space, Dirichlet (reflective) cavities and periodic cavities), we will give the full detail of the calculations for the continuum case and skip directly to the final results for periodic and Dirichlet cavities.

For the case of a field in free space (e.g. an open optical fibre or open transmission line) the field can be expanded in plane-wave modes as
\begin{equation}\label{eq:fieldfreespace}
%\phi(x,t) = \int_{-\infty}^{\infty} \frac{\d k\,}{\sqrt{2\omega_k}} \left(a_k e^{-\ii (\omega_kt-kx)}+a^\dagger_k e^{\ii (\omega_kt-kx)}\right)
\phi(x,t) = \int_{-\infty}^{\infty} \frac{\d k\,}{\sqrt{4\pi\omega_k}} \left(a^\dagger_k e^{\ii(\omega_kt-kx)}+\text{H.c.}\right)
\end{equation}
so that the interaction Hamiltonian becomes
\begin{equation*}
%H_I = i\lambda\chi(t)&{\mu}(t) \int_{-\infty}^{\infty}\d k\, \sqrt{\frac{\omega_k}{2}}\tilde{F}(k)\\
%&\left(-a_k e^{-\ii (\omega_kt-kx_a)}+a^\dagger_k e^{\ii (\omega_kt-kx_a)}\right),
\ii\lambda\chi(t){\mu}(t) \int_{-\infty}^{\infty}\d k\, \sqrt{\frac{\omega_k}{4\pi}}\tilde{F}(k)\left(a^\dagger_k e^{\ii(\omega_kt-kx_a)}-\text{H.c.}\right),
\end{equation*}
where  $\tilde{F}(k)=\int \d x\, f(x)e^{\ii kx}$ is the Fourier transform of the atomic spatial profile. We trace out the field to obtain the time-evolved state of the atom. The first order contribution to the time-evolved density matrix is traceless on the field for our initial state, therefore for an initial detector state given by
\begin{equation}
\ket{\psi}_A=a\ket{g}+\sqrt{1-a^2}\ket{e}
\end{equation}
and choosing a matrix representation such that
\begin{equation}
\ket{\psi}_A=\left(\begin{array}{c}a\\\sqrt{1-a^2}\end{array}\right),
\end{equation}
the second order contributions in \eqref{pertexp} are given by
\begin{equation*}
\Tr_{F}{(U^{(2)}\rho_i)} %&= -\lambda^2 \int_{-\infty}^{\infty}dt\int_{-\infty}^{t}dt' \chi(t)\chi(t') \int _{-\infty}^{\infty}\d k\, \frac{\omega_k}{2}\tilde{F}(k)^2 e^{-\ii \omega_k(t-t')} {\mu}(t){\mu}(t')\proj{\psi_A}{\psi_A}\\&
=\left( \begin{array}{cc}
a^2X_{++} & a\sqrt{1-a^2}X_{++}  \\
a\sqrt{1-a^2}X_{--} & (1-a^2)X_{--} \end{array} \right),
\end{equation*}
\begin{equation*}
\Tr_{F}{(U^{(1)}\rho_iU^{(1)\dagger})} %&= \lambda^2 \int_{-\infty}^{\infty}dt\int_{-\infty}^{\infty}dt' \chi(t)\chi(t') \int _{-\infty}^{\infty}\d k\, \frac{\omega_k}{2}\tilde{F}(k)^2 e^{\ii \omega_k(t-t')} {\mu}(t)\proj{\psi_A}{\psi_A}{\mu}(t')\\
=\left( \begin{array}{cc}
(1-a^2)J_{--} & a\sqrt{1-a^2}J_{-+}  \\
a\sqrt{1-a^2}J_{+-} & a^2J_{++} \end{array} \right).
\end{equation*}
Thus, the final state of the atom up to second order in perturbation theory is
\begin{widetext}
\begin{equation}
\label{eq:statefull}
\rho_A = \left( \begin{array}{cc}
a^2 & a\sqrt{1-a^2}  \\
a\sqrt{1-a^2} & 1-a^2 \end{array} \right) + \left( \begin{array}{cc}
(1-a^2)J_{--}+2a^2\Re{(X_{++})} & a\sqrt{1-a^2}(J_{-+}+X_{++}+X_{--}^*)  \\
a\sqrt{1-a^2}(J_{+-}+X_{++}^*+X_{--}) & a^2J_{++}+2(1-a^2)\Re{(X_{--})} \end{array} \right),
\end{equation}
where we define for $r,s\in\{+,-\}$
\begin{align}
\label{eq:XJcontinuum}
\begin{aligned}
X_{r,s} &=  -\lambda^2  \int _{-\infty}^{\infty}\d k\, \frac{\omega_k}{4\pi}\tilde{F}(k)^2 \int_{-\infty}^{\infty}\d t \int_{-\infty}^{t}\d t'\chi(t)\chi(t') e^{-\ii(\omega_k+r\Omega)t}e^{\ii(\omega_k+s\Omega)t'} \\
J_{r,s} &=  \lambda^2 \int _{-\infty}^{\infty}\d k\, \frac{\omega_k}{4\pi}\tilde{F}(k)^2 \int_{-\infty}^{\infty}\d t \int_{-\infty}^{\infty}\d t' \chi(t)\chi(t')  e^{\ii(\omega_k+r\Omega)t}e^{-\ii(\omega_k+s\Omega)t'}
\end{aligned}\,\,\mbox{(free space)}.
\end{align}
\end{widetext}

For atoms inside a cavity of length $L$, the field modes are no longer continuous but discrete. More specifically, for periodic boundary conditions (e.g. a closed optical fibre loop) we can make the following replacements
\begin{align}
k\rightarrow k_n=\frac{2\pi n}{L}, \ \omega_k\rightarrow \omega_n=\frac{2\pi |n|}{L}, \\ \int_{-\infty}^{\infty}\frac{\d k\,}{\sqrt{4\pi\omega_k}}\rightarrow \sum_{n=-\infty}^{\infty}\frac{1}{\sqrt{2\omega_nL}},
\end{align}
while for Dirichlet cavity  (e.g. reflective walls)
\begin{align}
k\rightarrow k_n=\frac{\pi n}{L}, \ \omega_k\rightarrow \omega_n=\frac{\pi n}{L}, \\ \int_{-\infty}^{\infty}\frac{\d k\,}{\sqrt{4\pi\omega_k}}\rightarrow \sum_{n=1}^{\infty}\frac{1}{\sqrt{\omega_nL}}.
\end{align}
Moreover, we make the physical assumption that the atom is much smaller than the size $L$ of the cavity, allowing us to simplify
\begin{align*}
\int_{-L/2}^{L/2}& \d x\,F(x-x_a)e^{\pm \ii k_nx}\\
&= e^{\pm \ii k_nx_a} \int_{-L/2-x_a}^{L/2-x_a}\d x\,F(x)e^{\pm \ii k_nx}\\
&\approx e^{\pm \ii k_nx_a} \int_{-\infty}^{\infty}F(x)e^{\pm \ii k_nx} =e^{\pm \ii k_nx_a}\tilde{F}(k_n)
\end{align*}
for a periodic cavity and similarly 
\begin{align*}
\int_{0}^{L}& \d x\,F(x-x_a)\sin(k_nx) \\
&\approx (e^{\ii k_nx_a}\tilde{F}(k_n)-e^{-\ii k_nx_a}\tilde{F}(-k_n))/2\ii\\ 
&= \tilde{F}(k_n)\sin(k_nx_a)
\end{align*}
for a Dirichlet cavity.

The form of the final state up to second order pertubation remains unchanged, and  $X$ and $J$ now take the following form
\begin{widetext}
\begin{equation}
\label{eq:Xrscavity}
X_{r,s} =
\begin{dcases}
-\lambda^2 \sum_{n=1}^{\infty} \frac{\omega_n}{L}\tilde{F}(k_n)^2 \int_{-\infty}^{\infty}\d t \int_{-\infty}^{t}\d t' \chi(t)\chi(t') e^{-\ii(\omega_n+r\Omega)t}e^{\ii(\omega_n+s\Omega)t'} & \mbox{(periodic)}\\
-\lambda^2  \sum_{n=1}^{\infty} \frac{\omega_n}{L}\tilde{F}(k_n)^2\sin^2(k_nx_a) \int_{-\infty}^{\infty}\d t \int_{-\infty}^{t}\d t' \chi(t)\chi(t') e^{-\ii(\omega_n+r\Omega)t}e^{\ii(\omega_n+s\Omega)t'} & \mbox{(Dirichlet)},
\end{dcases}
\end{equation}
\begin{equation}
\label{eq:Jrscavity}
J_{r,s} =
\begin{dcases}
\lambda^2 \sum_{n=1}^{\infty} \frac{\omega_n}{L}\tilde{F}(k_n)^2 \int_{-\infty}^{\infty}\d t \int_{-\infty}^{\infty}\d t' \chi(t)\chi(t')  e^{\ii(\omega_n+r\Omega)t}e^{-\ii(\omega_n+s\Omega)t'} & \mbox{(periodic)}\\
 \lambda^2 \sum_{n=1}^{\infty} \frac{\omega_n}{L}\tilde{F}(k_n)^2\sin^2(k_nx_a) \int_{-\infty}^{\infty}\d t \int_{-\infty}^{\infty}\d t' \chi(t)\chi(t')  e^{\ii(\omega_n+r\Omega)t}e^{-\ii(\omega_n+s\Omega)t'} & \mbox{(Dirichlet)}.
\end{dcases}
\end{equation}
\end{widetext}

\subsection{For comparison: The final atomic state under the single mode and rotating wave approximations}
When the coupling strength $\lambda$ is small, it is frequent in quantum optics to simplify the interaction Hamiltonian \eqref{eq:int} to the Jaynes-Cummings model where the single mode (SMA) and rotating wave (RWA) approximations are carried out when the atomic frequency is close to resonance with one of the cavity modes \cite{ScullyBook}. We discussed in the introduction that these two approximations yield non-relativistic models for light matter interaction. In this paper we will compare the predictions of extracted randomness of the fully relativistic calculation with the prediction of the usual RWA SMA prediction in the Jaynes-Cummings model.

One may wonder why in this paper we do not analyze the SMA and RWA separately, and that perhaps  only performing one of these two approximations could lead to valid results in the regimes that we are studying. However, we note that these two approximations are not independent, and in fact they derive from the same assumption: long evolution time as compared to the inverse of the atomic frequency gap. Moreover, in both approximations we neglect terms which are of the same order of magnitude, so it would be inconsistent to consider either of them individually (see Appendix A). Therefore, it makes sense to either do both approximations jointly (as we do in this section) or none of them (as we did in the previous section). %Indeed, if making one assumption is justified, then the terms it neglects are negligible, and so are the terms corresponding to the other assumption. On the other hand, if one assumption cannot be make, the terms neglected by this assumption are significant, and so must be the terms corresponding to the other assumption.

For the purpose of the comparison we suppose that the atom is on resonance with the $m^{th}$ ($m>0$) mode of the cavity, namely $\Omega=2\pi m/L$ for periodic cavity and $\Omega=\pi m/L$ for Dirichlet cavity (which can be obtained by controlling the cavity's length). With $b_m=\frac{1}{\sqrt{2}}(a_{_{m}}e^{\ii k_mx_a}+a_{_{\!-\!m}}e^{-\ii k_mx_a})$, $k_m=\Omega$ as the resonant standing wave mode of the periodic cavity, the interaction Hamiltonian under RWA and SMA becomes
\begin{equation*}
H_I=\ii\lambda\chi(t)\sqrt{\frac{\Omega}{L}}\tilde{F}(k_m) \left(-\sigma_+b_m +\sigma_-b^\dagger_m \right)
\end{equation*}
for a periodic cavity, and
\begin{equation*}
H_I=i\lambda\chi(t)\sqrt{\frac{\Omega}{L}} \tilde{F}(k_m)\sin(k_mx_a) \left(-\sigma_+a_m+\sigma_-a^\dagger_m\right)
\end{equation*}
for a Dirichlet cavity. This model can be solved exactly for all times, yielding the final state
\begin{equation*}
\rho_A^m = 
\left( \begin{array}{cc}
a^2 +(1-a^2)\sin^2\left(\Theta\right) & a\sqrt{1-a^2}\cos\left(\Theta\right)  \\
a\sqrt{1-a^2}\cos\left(\Theta\right) & (1-a^2)\cos^2\left(\Theta\right) \end{array} \right),
\end{equation*}
where
\begin{equation*}
\Theta=
\begin{dcases}
\frac{\lambda\Omega}{\sqrt{2\pi m}}\tilde{F}(\Omega)T & \mbox{(periodic)}\\
\frac{\lambda\Omega}{\sqrt{\pi m}}\tilde{F}(\Omega)\sin(\Omega x_a)T & \mbox{(Dirichlet)}
\end{dcases}
\end{equation*}
and therefore the predictions of the SMA-RWA Jaynes-Cummings model can be compared with the fully relativistic model within the perturbative regime. More precisely, we use the following second order expansion of the previous final state
\begin{equation}
\label{eq:stateRWA}
\rho_A^m = 
\left( \begin{array}{cc}
a^2 +(1-a^2)\Theta^2 & a\sqrt{1-a^2}\left(1-\frac{\Theta^2}{2}\right)  \\
a\sqrt{1-a^2}\left(1-\frac{\Theta^2}{2}\right) & (1-a^2)(1-\Theta^2) \end{array} \right),
\end{equation}
in the comparison.

\section{Simulation results}

\subsection{\label{sec:concrete}The concrete simulation model}
Between the instant the atom is prepared in the state $\ket{\psi}_A$ and its measurement, the atom interacts with the field for a duration $T$ in a manner that can be captured by the sharp switching function
\begin{equation}
\chi(t) = \left\{ \begin{array}{ll}
         0 & \mbox{if $t \leq 0$}\\
         1 & \mbox{if $0 < t \leq T$}\\
         0 & \mbox{if $t > T$}.\end{array} \right.
\end{equation}
We assume that the atom has the following simple spatial profile:
\begin{equation}
F(x) = \frac{1}{\sigma\sqrt{\pi}}e^{-x^2/\sigma^2}, \ \ \tilde{F}(k)= e^{-\sigma^2k^2}.
\end{equation}
where $\sigma$ gives the characteristic lenghtscale of the atomic species. Under these conditions, we have
\begin{widetext}
\begin{align*}
X_{\pm,\pm} &=
\begin{dcases}
 -\lambda^2  \int _{0}^{\infty}\d k\, \frac{k}{2\pi}\tilde{F}(k)^2 \left[\frac{T}{\ii ( k\pm\Omega)}-\frac{1}{(k\pm\Omega)^2}(e^{-\ii (k\pm\Omega)T}-1)\right]& \mbox{(free space)}\\
 -\lambda^2 \sum_{n=1}^{\infty} \frac{2\pi n}{L^2}\tilde{F}\left(\frac{2\pi n}{L}\right)^2 \left[\frac{T}{\ii ( \frac{2\pi n}{L}\pm\Omega)}-\frac{1}{(\frac{2\pi n}{L}\pm\Omega)^2}(e^{-\ii (\frac{2\pi n}{L}\pm\Omega)T}-1)\right]& \mbox{(periodic)}\\
-\lambda^2  \sum_{n=1}^{\infty} \frac{\pi n}{L^2}\tilde{F}\left(\frac{\pi n}{L}\right)^2\sin^2\left(\frac{\pi nx_a}{L}\right)\left[\frac{T}{\ii ( \frac{\pi n}{L}\pm\Omega)}-\frac{1}{(\frac{\pi n}{L}\pm\Omega)^2}(e^{-\ii (\frac{\pi n}{L}\pm\Omega)T}-1)\right]& \mbox{(Dirichlet)},
\end{dcases}\\
J_{\pm,\pm} &=  
\begin{dcases}
\lambda^2 \int _{0}^{\infty}\d k\, \frac{2\tilde{F}(k)^2k}{\pi(k\pm\Omega)^2}\sin^2\left(\frac{(k\pm\Omega)T}{2}\right) & \mbox{(free space)}\\
\lambda^2 \sum_{n=1}^{\infty} \frac{8\pi n}{L^2}\tilde{F}\left(\frac{2\pi n}{L}\right)^2 \frac{1}{(\frac{2\pi n}{L}\pm\Omega)^2}\sin^2\left(\frac{(\frac{2\pi n}{L}\pm\Omega)T}{2}\right) & \mbox{(periodic)}\\
\lambda^2  \sum_{n=1}^{\infty} \frac{4\pi n}{L^2}\tilde{F}\left(\frac{\pi n}{L}\right)^2\sin^2\left(\frac{\pi nx_a}{L}\right) \frac{1}{(\frac{\pi n}{L}\pm\Omega)^2}\sin^2\left(\frac{(\frac{\pi n}{L}\pm\Omega)T}{2}\right)& \mbox{(Dirichlet)},
\end{dcases}\\
J_{\pm,\mp} &= 
\begin{dcases}
\lambda^2 \int _{0}^{\infty}\d k\, \frac{\tilde{F}(k)^2k}{2\pi(k^2-\Omega^2)}\left[1+e^{\pm2\ii \Omega T}-2\cos(kT)e^{\pm \ii \Omega T}\right] & \mbox{(free space)}\\
\lambda^2\sum_{n=1}^{\infty}  \frac{2\pi n}{L^2} \tilde{F}\left(\frac{2\pi n}{L}\right)^2 \frac{1}{(\frac{2\pi n}{L})^2-\Omega^2} \left[1+e^{\pm2\ii \Omega T}-2\cos\left(\frac{2\pi n}{L}T\right)e^{\pm \ii \Omega T}\right] & \mbox{(periodic)}\\
\lambda^2  \sum_{n=1}^{\infty} \frac{\pi n}{L^2}\tilde{F}\left(\frac{\pi n}{L}\right)^2\sin^2\left(\frac{\pi nx_a}{L}\right)  \frac{1}{(\frac{\pi n}{L})^2-\Omega^2} \left[1+e^{\pm2\ii \Omega T}-2\cos\left(\frac{\pi n}{L}T\right)e^{\pm \ii \Omega T}\right]& \mbox{(Dirichlet)},\\
\end{dcases}
\end{align*}
\end{widetext}
where the notation $X_{\pm,\pm}$ refers to either the upper $X_{+,+}$ or the lower $X_{-,-}$ combination of signs to be taken on the right hand side (the same for the others). Given these expressions, the final state after the interaction~\eqref{eq:statefull} can be numerically approximated with high accuracy by performing the numerical integration or numerical summation up to a cutoff $N_c/\sigma$. Here, we normalize the numerical cutoff $N_c$ by the atomic's size $\sigma$. Since the Fourier transform of the spatial profile is a Gaussian centered at zero with standard deviation proportional to $\sigma^{-1}$, taking $N_c\simeq 6$ already gives an extremely precise numerical approximation, independently of the atom's size.

% For an atom inside a cavity, if the atom's enegery gap is on resonance with one of the cavity modes, namely $\Omega=\frac{2\pi |n|}{L}$ for periodic cavity and $\Omega=\frac{\pi n}{L}$, then the contribution of this mode cannot be calculated directly by substitution into the summation above for this term will be indeterminate. We must calculate it from the equations~\eqref{eq:Xrscavity} and \eqref{eq:Jrscavity}, by doing the time integrals directly for the on resonant mode. The result of this excercise are the following contributions
% \begin{align*}
% X^n_{-,-} &= 
% \begin{dcases}
% \frac{-\lambda^2\Omega}{2L}\tilde{F}(\Omega)^2T^2 & \mbox{(periodic)},\\
% \frac{-\lambda^2\Omega}{2L}\tilde{F}(\Omega)^2\sin^2(\Omega x_a)T^2 & \mbox{(Dirichlet)},
% \end{dcases}\\
% J^n_{-,-} &= 
% \begin{dcases}
% \frac{\lambda^2\Omega}{L}\tilde{F}(\Omega)^2T^2 & \mbox{(periodic)},\\
% \frac{\lambda^2\Omega}{L}\tilde{F}(\Omega)^2\sin^2(\Omega x_a)T^2 & \mbox{(Dirichlet)},
% \end{dcases}\\
% J^n_{+,-} &= 
% \begin{dcases}
% \frac{-i\lambda^2}{2L}\tilde{F}(\Omega)^2T(e^{2\ii \Omega T}-1) & \mbox{(periodic)},\\
% \frac{-i\lambda^2}{2L}\tilde{F}(\Omega)^2\sin^2(\Omega x_a)T(e^{2\ii \Omega T}-1) & \mbox{(Dirichlet)}.
% \end{dcases}
% \end{align*}
% In other words, the cavity summations can be numerically computed by direct substitution of all the parameters into the summands where resonant does not occur, and for those $n$ where resonance happens, given by the above $X^n$ and $J^n$.

\subsection{Randomness certification in free space}
\begin{figure}[h]
\includegraphics[width=0.50\textwidth]{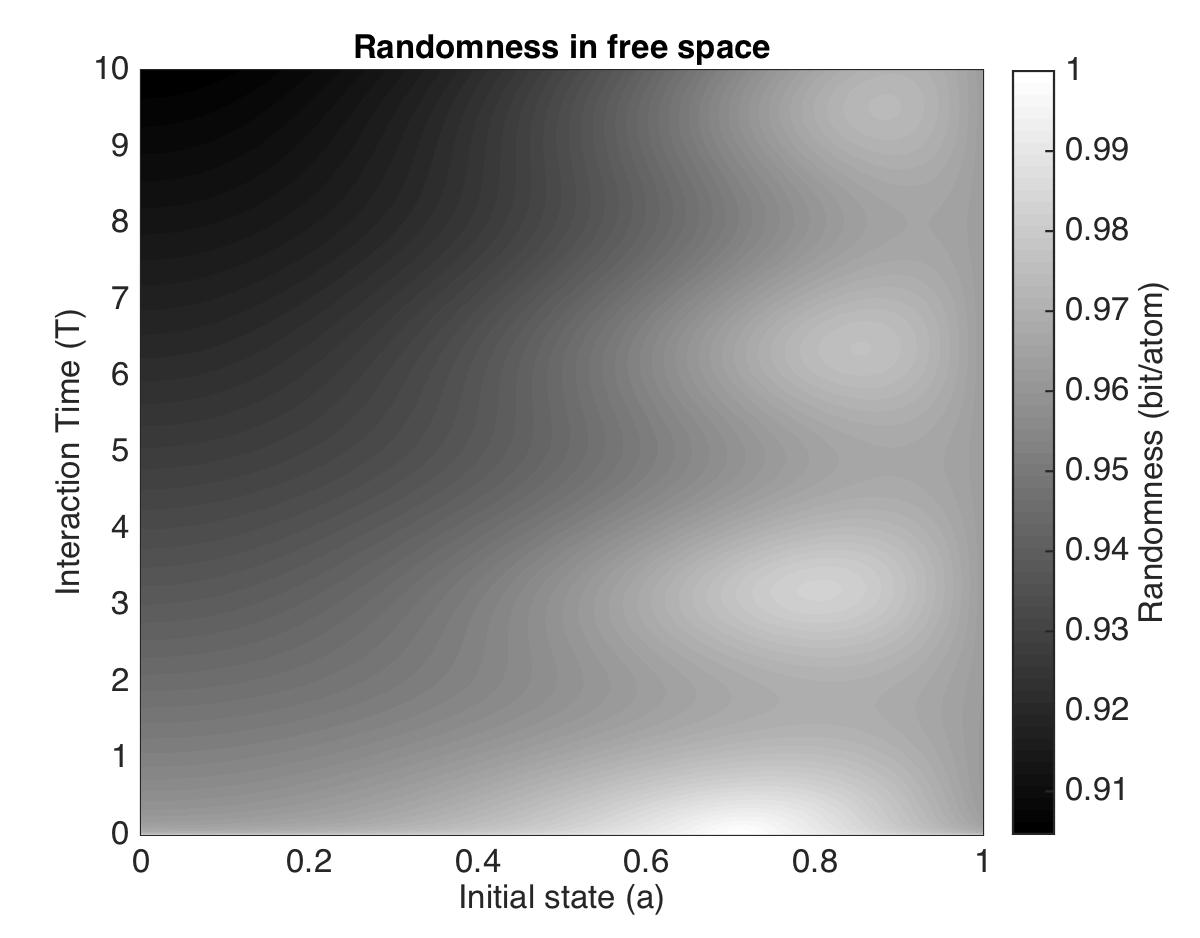}
\caption{\label{fig:RvsStateTime}Randomness in the free space scenario for different initial states at different measurement time after preparation, with chosen parameters $\lambda =0.01, \sigma =0.001, \Omega =1$ and $N_c=6$.}
\end{figure}

From the final states computed in the previous sections, one can compute the number of random bits that can be extracted per atom measurement using Eq.\eqref{eq:randopt}. In Fig.~\ref{fig:RvsStateTime} we report the result of this computation for the free field case (i.e. using Eq.\eqref{eq:statefull},\eqref{eq:XJcontinuum}).

\begin{figure*}[t]
\includegraphics[width=0.32\textwidth]{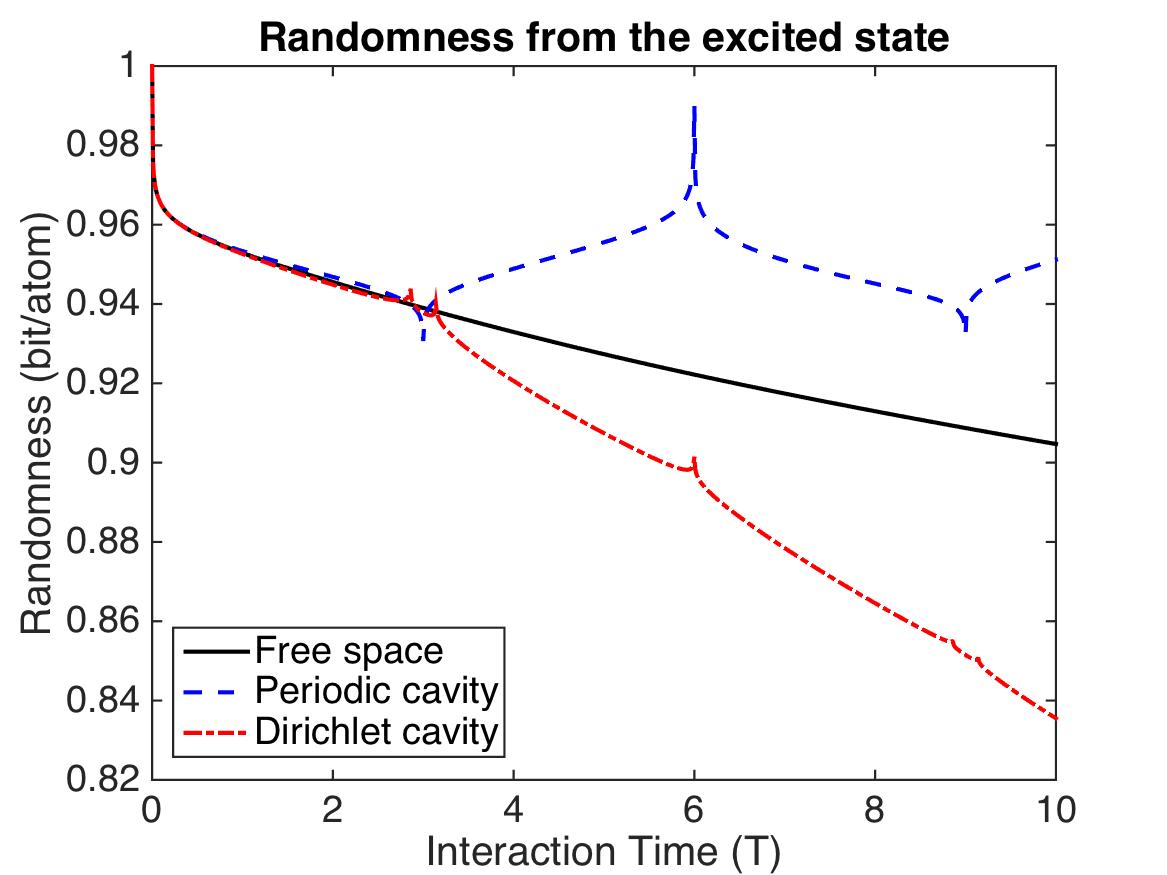}
\includegraphics[width=0.32\textwidth]{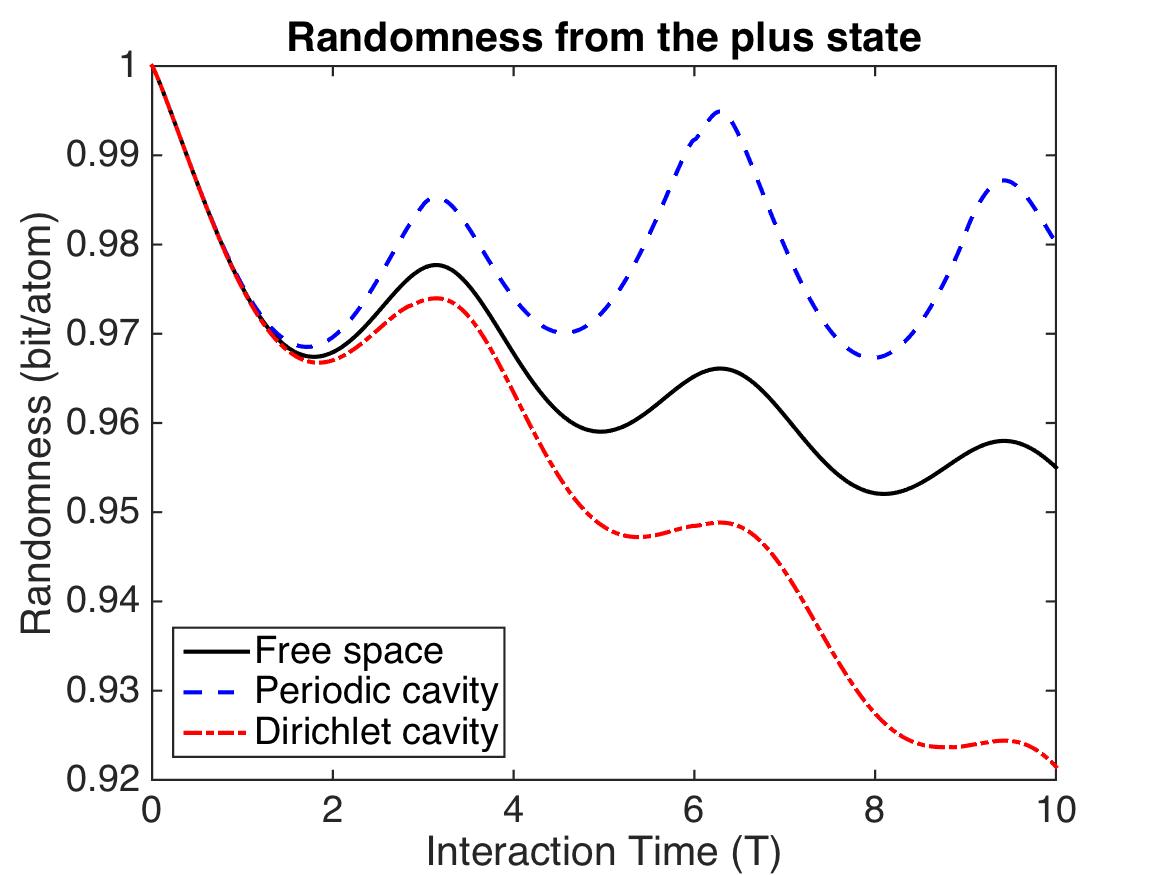}
\includegraphics[width=0.32\textwidth]{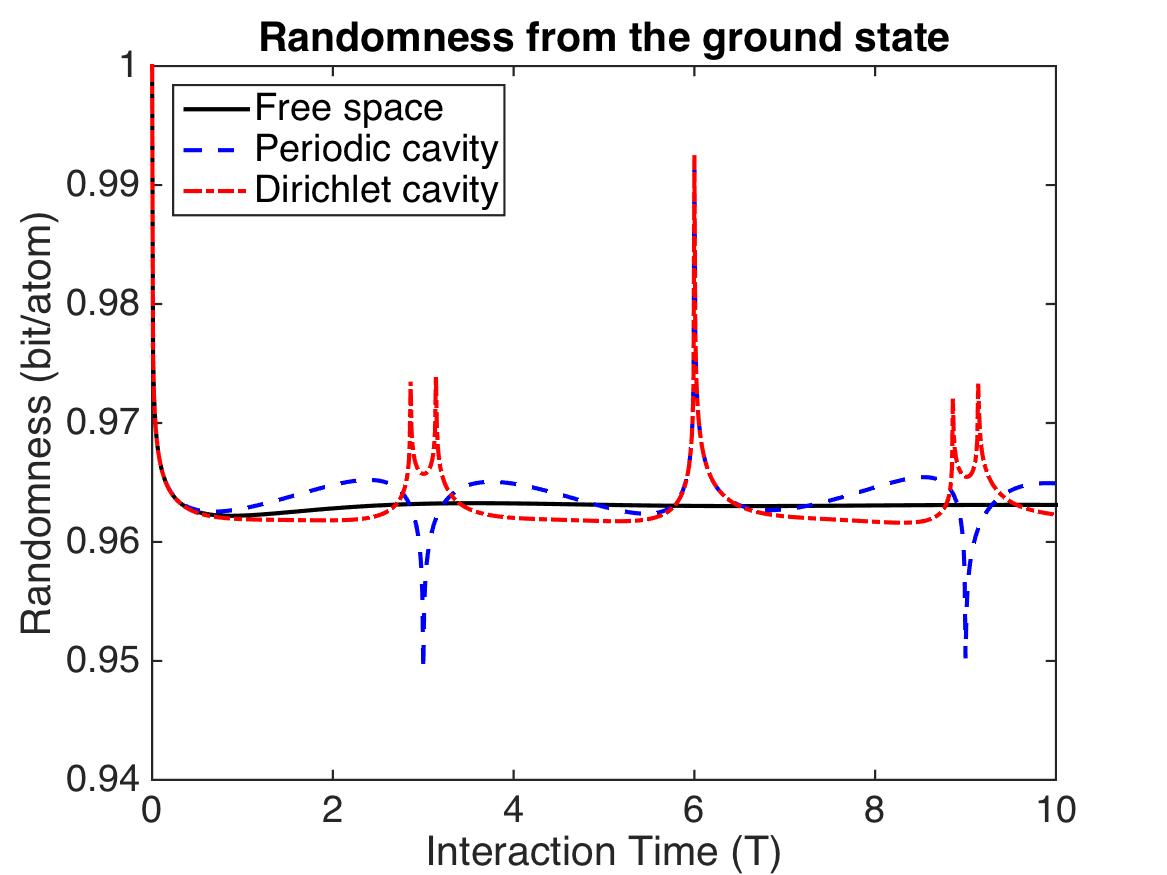}
\caption{\label{fig:RvsTime}Randomness for (from left to right) $\ket{e},\ket{+}$, and $\ket{g}$ states at different measurement time after preparation, with chosen parameters $\lambda =0.01, \sigma =0.001, \Omega =1$,  $N_c=6$, $L=3$ and $x_a=\pi L/6$. As the length $L$ of the cavities increases, we observed that the cavity curves (blue dashed and red dot-dashed)  converge to the free space curve (black solid). Note that the position of the atom in Dirichlet cavity is chosen to be the default position $x_a=\pi L/6$ which is roughly in the middle of the cavity.  The peaks correspond to the light-crossing time of the cavity: The perturbation caused by the introduction of the atom returns to the atom after scattering with the Dirichlet/Periodic boundaries of the cavities.}
\end{figure*}

A first clear observation from Fig.~\ref{fig:RvsStateTime} is that for most initial states, the randomness rate quickly decreases from 1 as soon as $T>0$ (see also solid line in Fig.~\ref{fig:RvsTime}). This shows that high-frequency terms play an important role in the evolution of the state for short times, and therefore they cannot be neglected.

One expected result that is verified in Fig.~\ref{fig:RvsStateTime} is that preparing the atom in an excited state ($a=0$) always gives less randomness, at all interaction times within the limits of pertubation theory, than preparing it in the ground state. The reason for this behavior is clear: an atom in the excited state can be de-excited by the rotating wave terms in the interaction Hamiltonian with an elevated probability by emitting real field quanta. For short times, these field quanta are therefore correlated with the state of the atom, giving away information about the atomic state to an adversary having access to the quantum field. Conversely, an atom in the ground state ($a=1$) can only be correlated with the field via the counter-rotating part of the Hamiltonian. Even though in that case the atom also gets correlated with the field through vacuum fluctuations, the excited state is always less secure. This behavior is indeed expected from the non-relativistic intuition that an excited atom may emit a photon that an adversary can capture and learn about the state of the atom.

However, Fig.~\ref{fig:RvsStateTime} also reveals a less intuitive effect. Namely, the ground state is not always the optimal state to extract randomness: depending on the other parameters, most notably the lag time between preparation and measurement $T$, it may be better to prepare the atom in a superposition of ground and excited state (see also solid line in Fig.~\ref{fig:RvsTime}).

It is clear that the ground state cannot be fully secured, because it is not an eigenstate of the interaction Hamiltonian. Therefore the interaction introduces correlations between the atom and the field even when starting from the ground state. These correlations can later be used by an adversary (that does not need to be in light-contact with the first atom) to learn about the result of a measurement on the original atom. An example of how an adversary can gain information about the outcome of measurements even without receiving any energy from it is the `quantum collect calling' (virtual-photon mediated timelike communication)~\cite{jonsson_quantum_2014,Jonsson2015,ComsoQCC}.

It is worth noting that the randomness certified when starting from the ground state, after rapidly decreasing, seems to attain an asymptotic value (see Fig~\ref{fig:RvsTime}). While it is out of the scope of the present paper, it may constitute an interesting follow-up work to check whether this is still the case in the long time regimes, or whether non-perturbative effects may still significantly change the purity of the reduced state of the atom $\rho_A$ for long times.

\begin{figure}[h]
\includegraphics[width=0.5\textwidth]{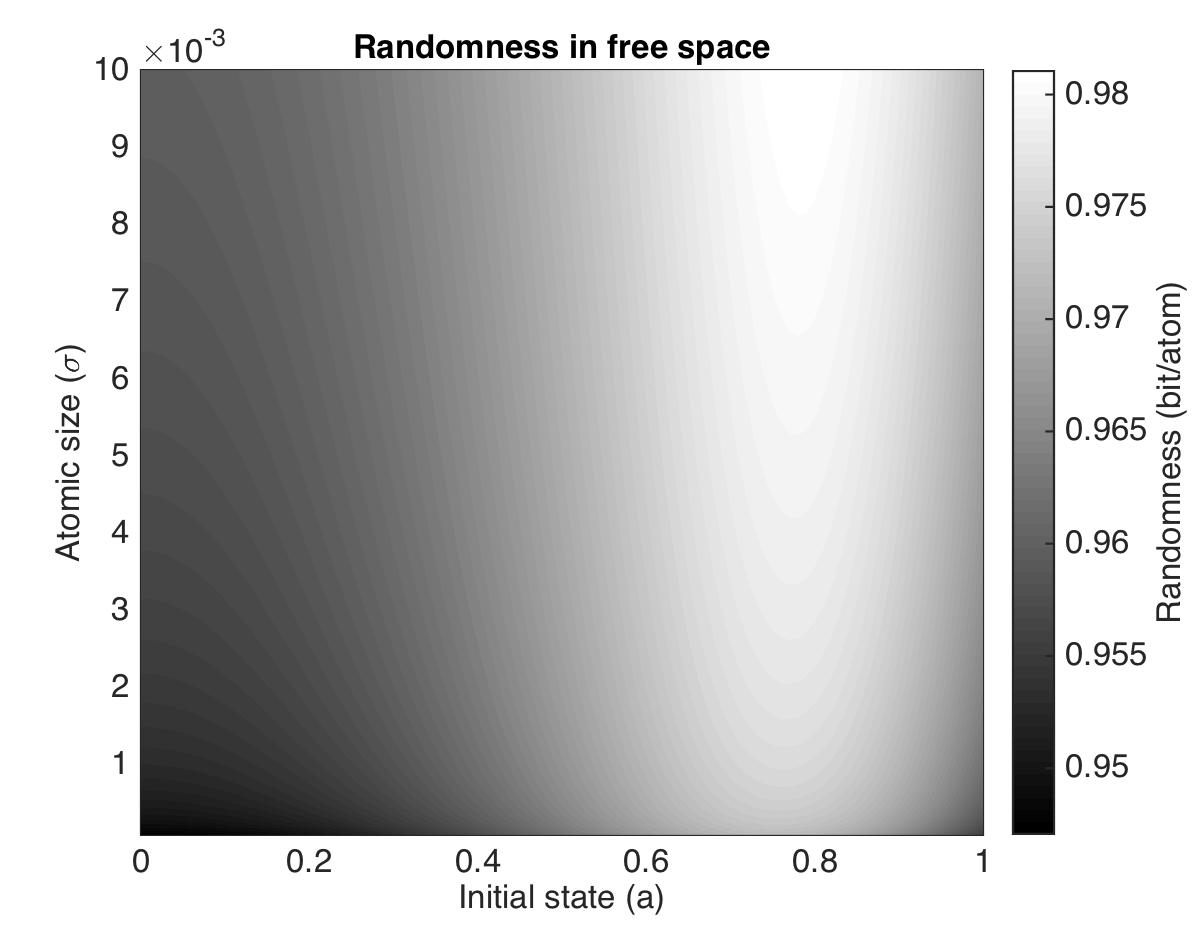}
\caption{\label{fig:RvsStateSize}Randomness for different initial states and atomic sizes, with chosen parameters $\lambda=0.01, \Omega=1, T=1$ and $N_c=6$.}
\end{figure}
In Fig.~\ref{fig:RvsStateSize}, we study the effect of the atomic size on the certified randomness. One observes that more randomness is certified in presence of large atoms. The reason for this is that the bigger the atom gets the less the atom couples to the highest frequencies of the field, so the less the initial state is affected. Notice that the single mode approximation is recovered here  when the atom is taken to be infinitely large and, thus, couples to a single frequency. This case, of course, breaks the relativistic approach (the single mode approximation strongly violates causality \cite{jonsson_quantum_2014}) which is not surprising since the atom sees the field at all points in space at the same time. In this case, the amount of certified randomness is essentially uniform over all states. %The peaks in Fig. \ref{fig:RvsTime} correspond to the light-crossing time of the cavity (the perturbation caused by the switching of the interaction bounces back on the boundary of the cavity and returns to the atom).

\begin{figure*}[t]
\includegraphics[width=0.32\textwidth]{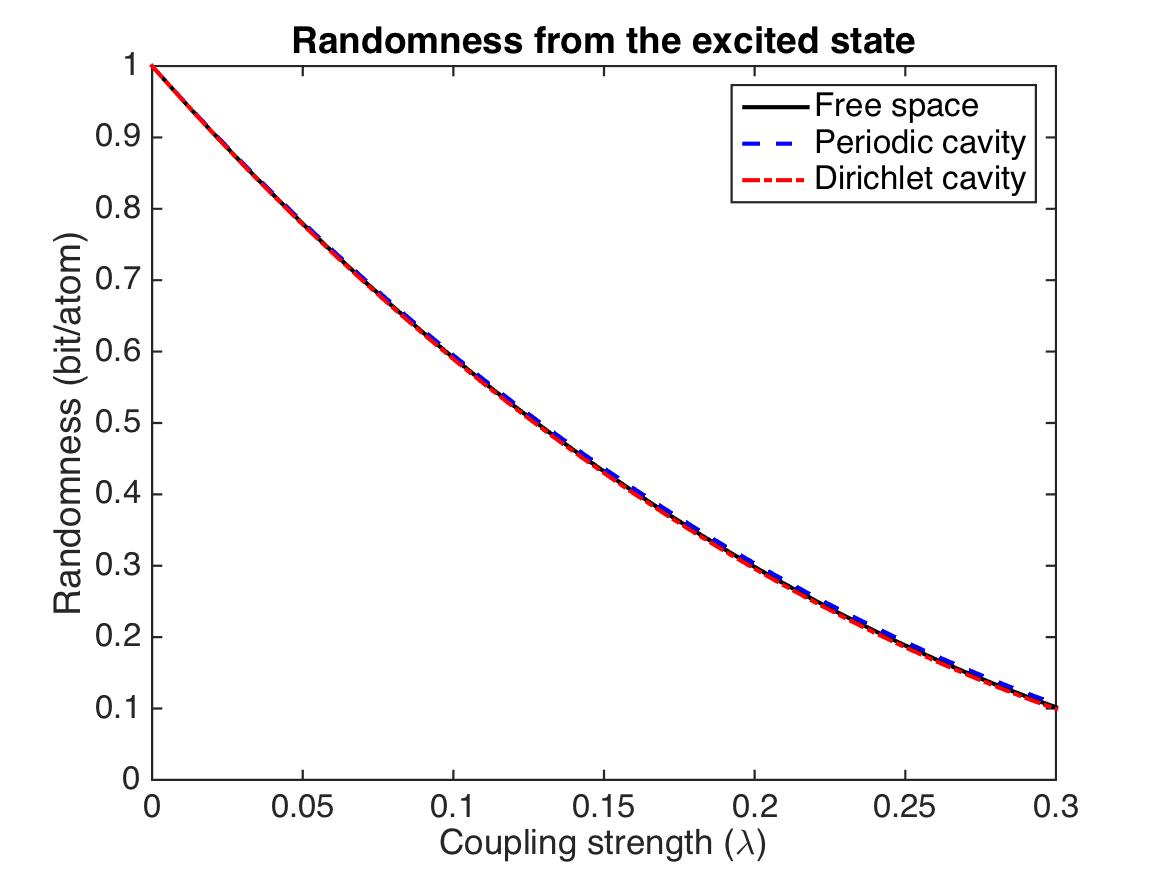}
\includegraphics[width=0.32\textwidth]{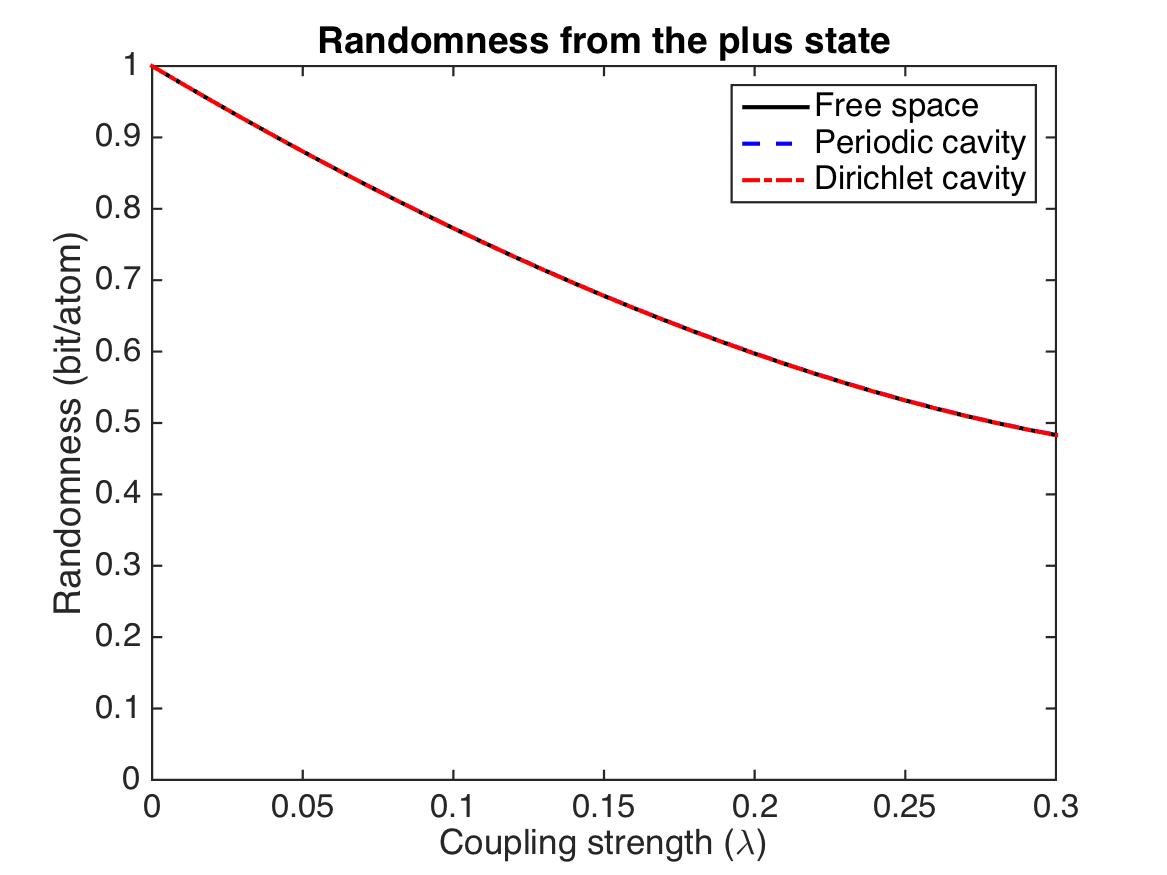}
\includegraphics[width=0.32\textwidth]{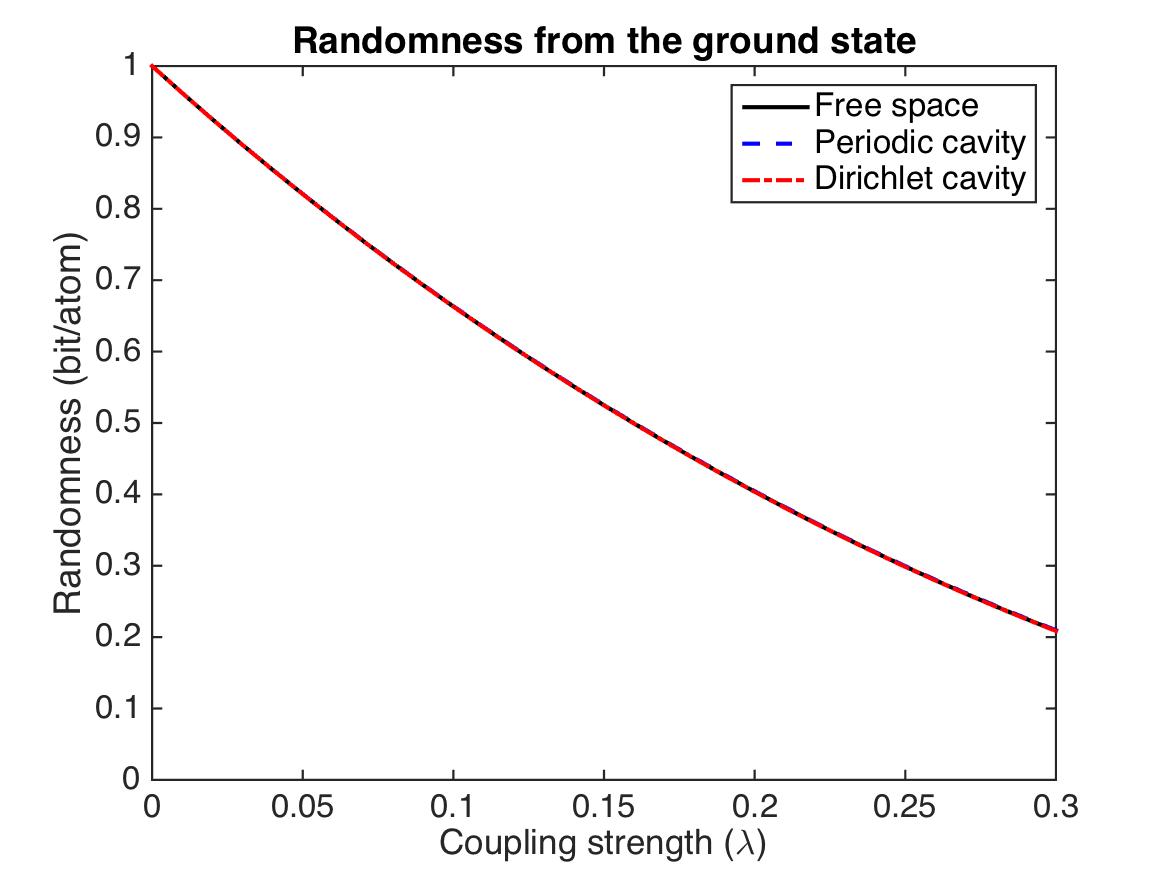}
\caption{\label{fig:RvsLambda}Randomness vs coupling strength, with chosen parameters  $\sigma=0.001, \Omega=1, T=1$, $N_c=6$, $L=3$ and $x_a=\pi L/6$.}
\end{figure*}

%This is remarkable given that short-time effects (relativistic) such as entanglement harvesting \cite{Reznik2003,Valentini1991,Reznik2005} \textbf{citation?} or simply the \textbf{?} are usually impacting \textbf{? [complete the sentence]}. This means that for a small coupling strength, relativistic short-time effects have a stronger influence on randomness than other effects commonly seen is relativistic quantum information {\bf we need to elaborate?}.

\subsection{Randomness extraction in cavity}
Atoms inside cavities are a more realistic experimental scenario compared to atoms in free space \cite{nature,Sergelecture,nature3,nature1,Haroche1,singlephotons}. There are two main differences when atoms are put inside a cavity. Firstly, the cavity only supports a countable infinite number of modes, as opposed to the continuously many modes in free space. Secondly, although there are fewer modes to interact, the interaction may be made stronger than in free space \cite{Blais04,Wallraff04,Chiorescu04,Peropadre10}.

The resulting effect of these two differences on the guessing probability is presented in FIG.~\ref{fig:RvsTime}. The periodic and Dirichlet curves in this graph were obtained by computing Eq.~\eqref{eq:randopt} with Eq.~\eqref{eq:statefull},\eqref{eq:Xrscavity},\eqref{eq:Jrscavity}. Notice that the length of the cavity in this case is three orders of magnitude larger than the  size of the atom, so our physical assumption of a small atom in a large cavity is met. This figure also compares the randomness rate with respect to the one obtained in the free field case. The peaks in Fig. \ref{fig:RvsTime} correspond to the light-crossing time of the cavity (the perturbation caused by the switching of the interaction bounces back on the boundary of the cavity and returns to the atom).

\begin{figure*}[t]
\includegraphics[width=0.32\textwidth]{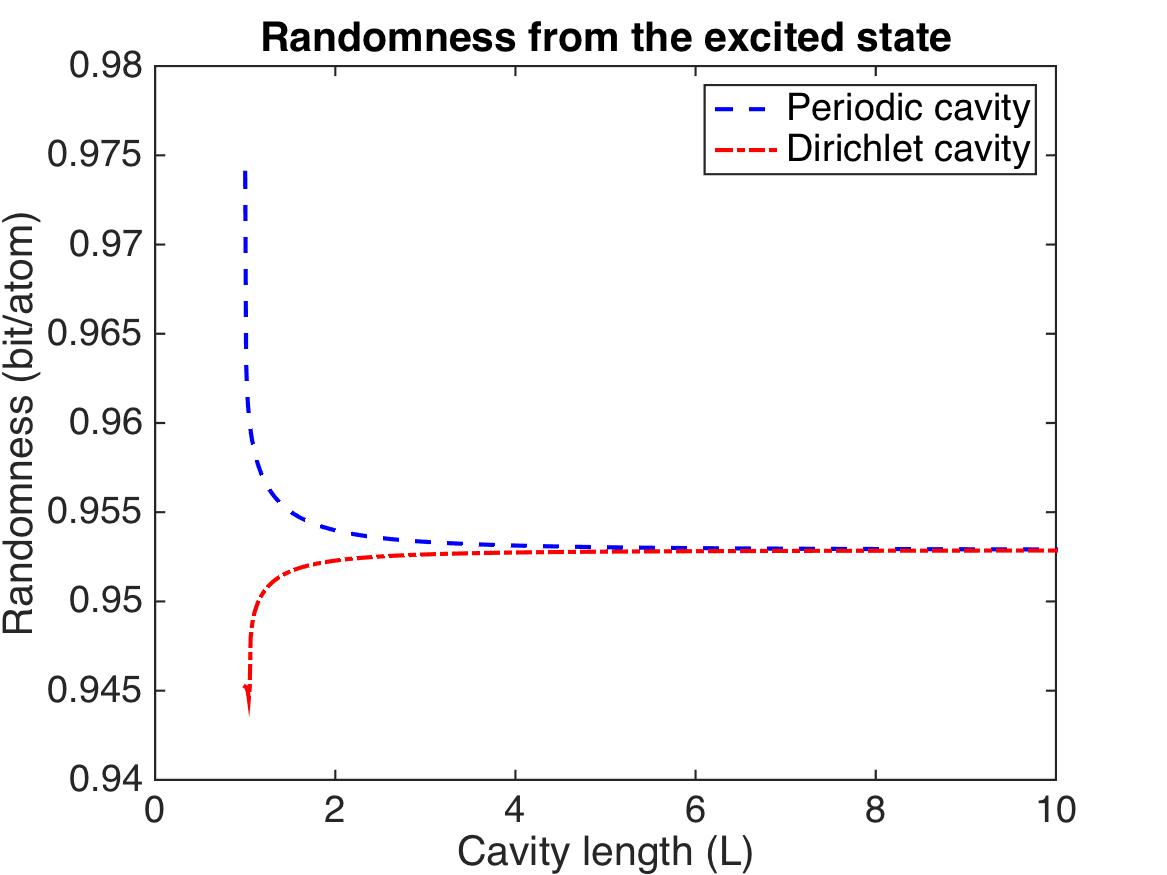}
\includegraphics[width=0.32\textwidth]{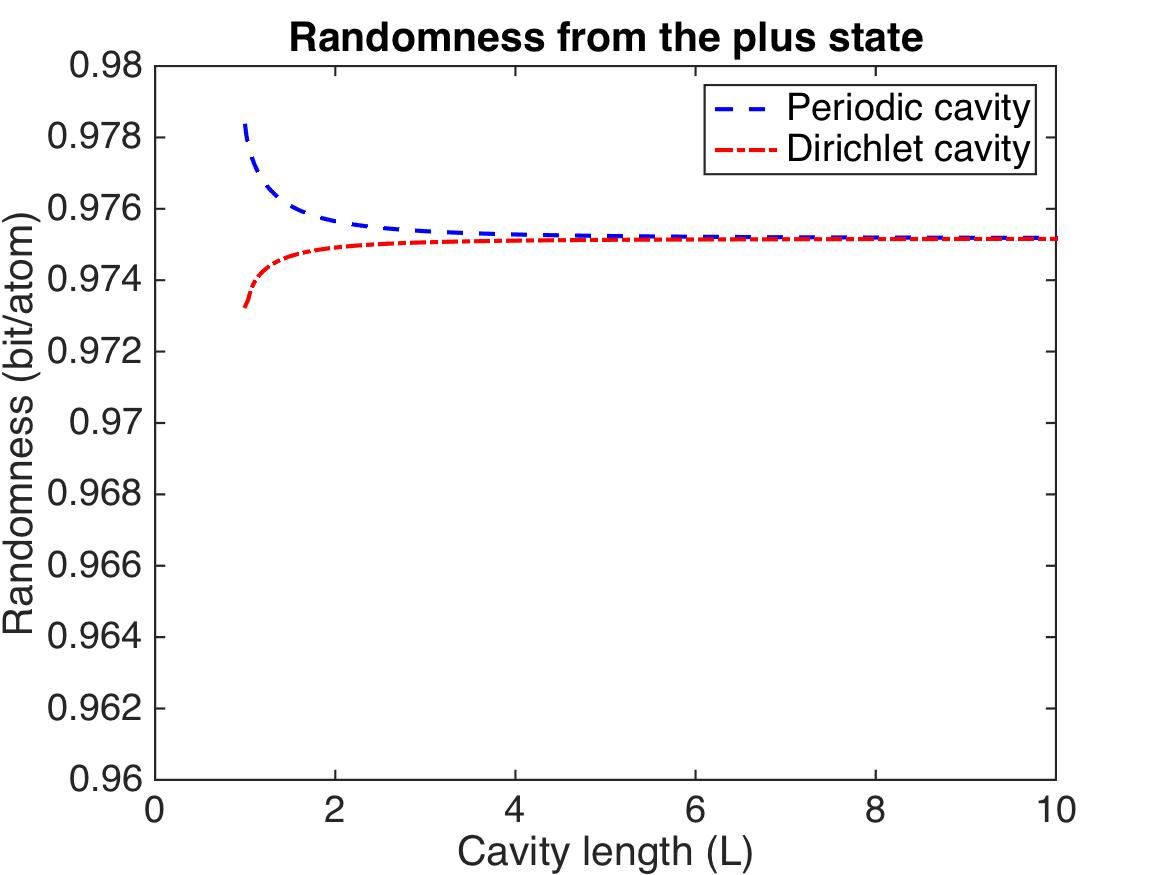}
\includegraphics[width=0.32\textwidth]{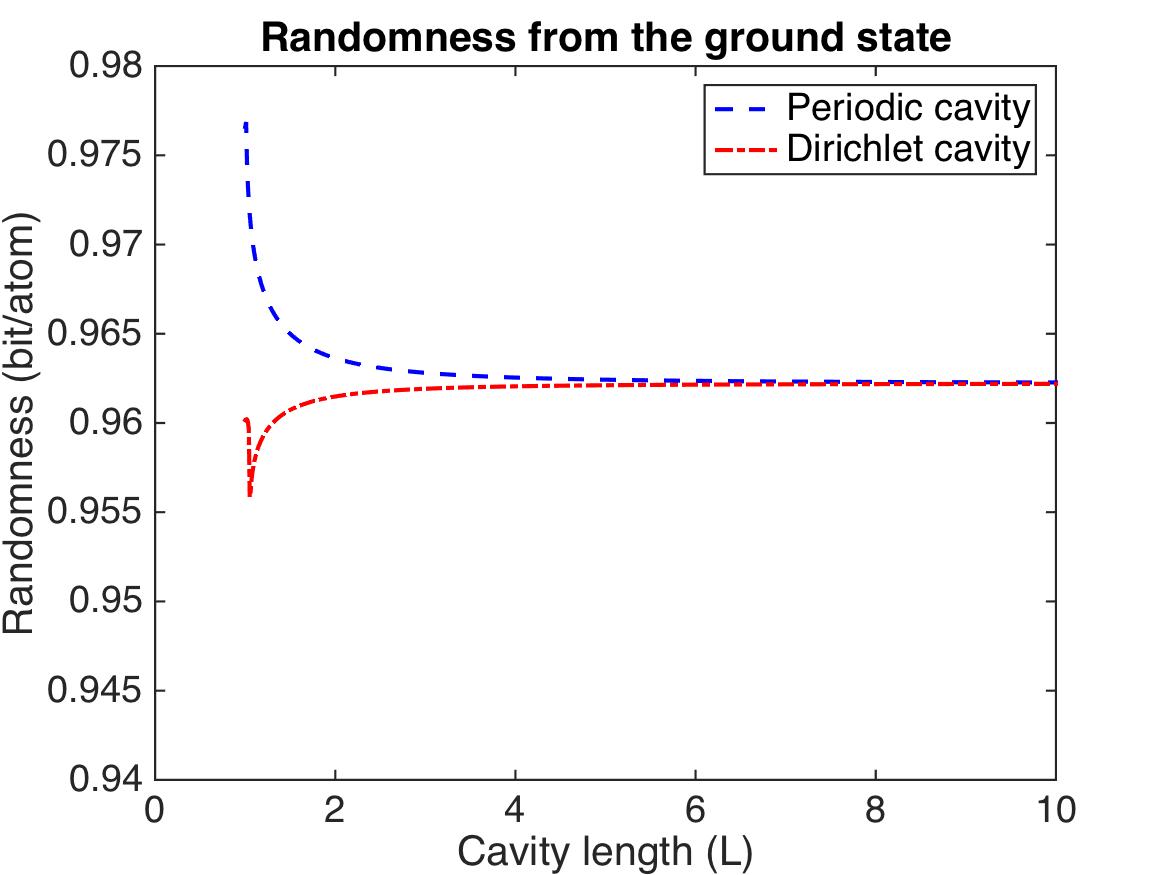}
\caption{\label{fig:RvsL}Randomness for various cavity's lengths assuming the atom was prepared in the state $\ket{e}$,$\ket{+}$ or $\ket{g}$, with chosen paramters  $\lambda=0.01, \sigma=0.001, \Omega=1, T=1, N_c=6$ and $x_a=\pi L/6$. Dirichlet (periodic) cavity randomness converges \textit{up} (\textit{down}) to free space randomness.}
\end{figure*}

One would expect that for larger and larger cavities, the two cavity results converge to the free space one. This is verified in Fig.~\ref{fig:RvsL}.

\begin{figure}[h]
\includegraphics[width=0.5\textwidth]{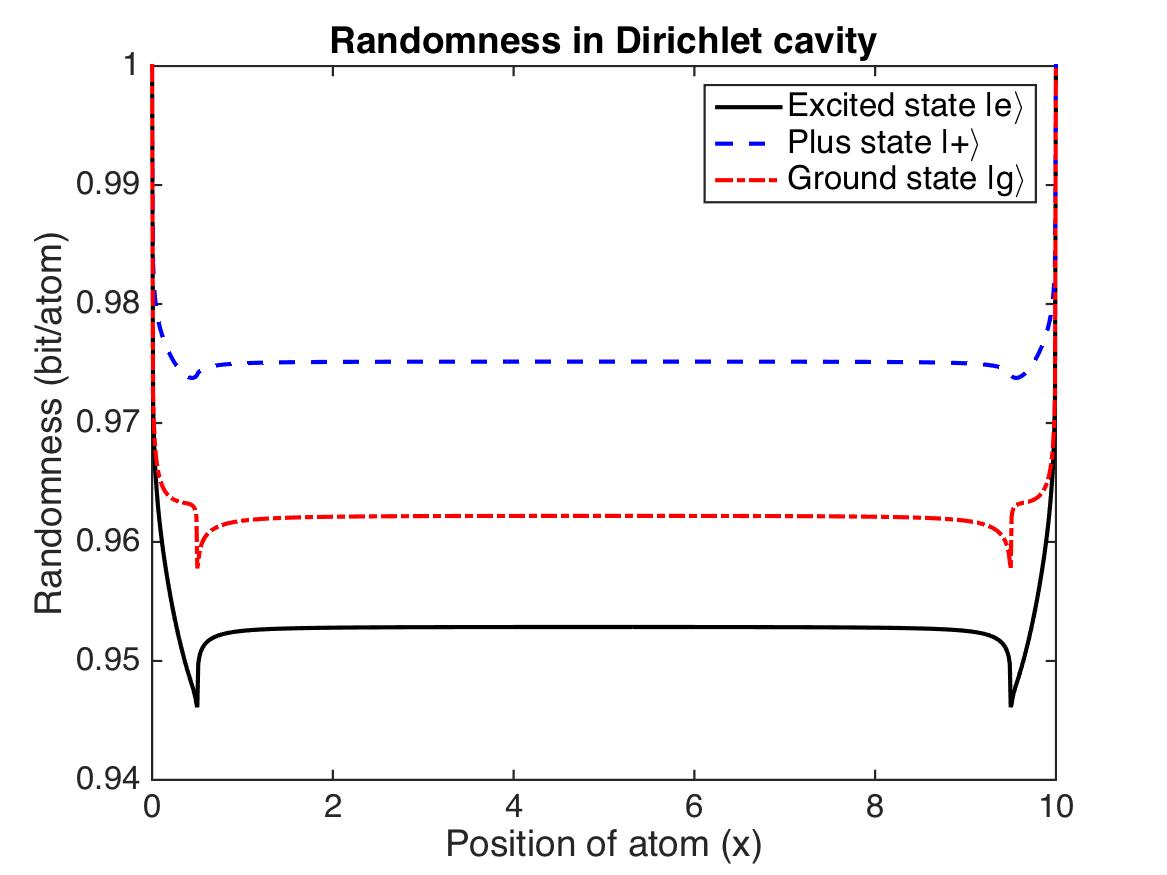}
\caption{\label{fig:RvsX}Randomness vs position of atom in Dirichlet cavity, with chosen parameters  $\lambda=0.01,\sigma=0.001,\Omega=1,T=1,L=10$ and $N_c=6$.}
\end{figure}

Also, in a Dirichlet cavity, the randomness output depends explicitly on the position of the atom, unlike in a periodic cavity. Fig.~\ref{fig:RvsX} shows that this dependence is negligible.

Finally, in Fig.~\ref{fig:RvsLambda} we analyse the role of the coupling strength on the randomness rate. We observe that the behavior is the same for all the boundary condition scenarios. This can be understood because we know from equation~\eqref{eq:randopt} that the min-entropy only depends on the \textit{purity} of the final state $\Tr{\rho_A^2}$, and from \eqref{eq:int}, \eqref{pertexp} and \eqref{afterfour}, we see that the purity of the atomic state scales as $1-\lambda^2$ for any set of boundary conditions.

\subsection{Comparison with the rotating wave approximation}
In order to compare the above results with predictions of the RWA model, we introduce the difference ratio
\begin{equation}
R=\frac{H_{\min}^\textrm{RWA}-H_{\min}^\textrm{full}}{H_{\min}^\textrm{RWA}},
\end{equation}
where $H_{\min}^\textrm{full}$ is the randomness computed according the the method presented in the precedent paragraphs, i.e. from the state~\eqref{eq:statefull} with the terms described in Section~\ref{sec:concrete}, and $H_{\min}^\textrm{RWA}$ stands for the randomness computed directly from the state~\eqref{eq:stateRWA}.

Since the RWA randomness is computed under the assumption that the atom is on resonant with some $m^\textup{th}$ mode of the cavity, throughout this section, the length of the cavity is always fixed based on the chosen resonant mode $m$ according to $L=2\pi m/\Omega$ for periodic cavity and $L=\pi m/\Omega$ for Dirichlet cavity. Note that for the sake of numerical simulation, $m$ cannot be chosen too small relative to $\sigma$ because this violates the assumption of a relatively big cavity with respect to the atom's size which we have made before.

\begin{figure*}[t]
\includegraphics[width=0.49\textwidth]{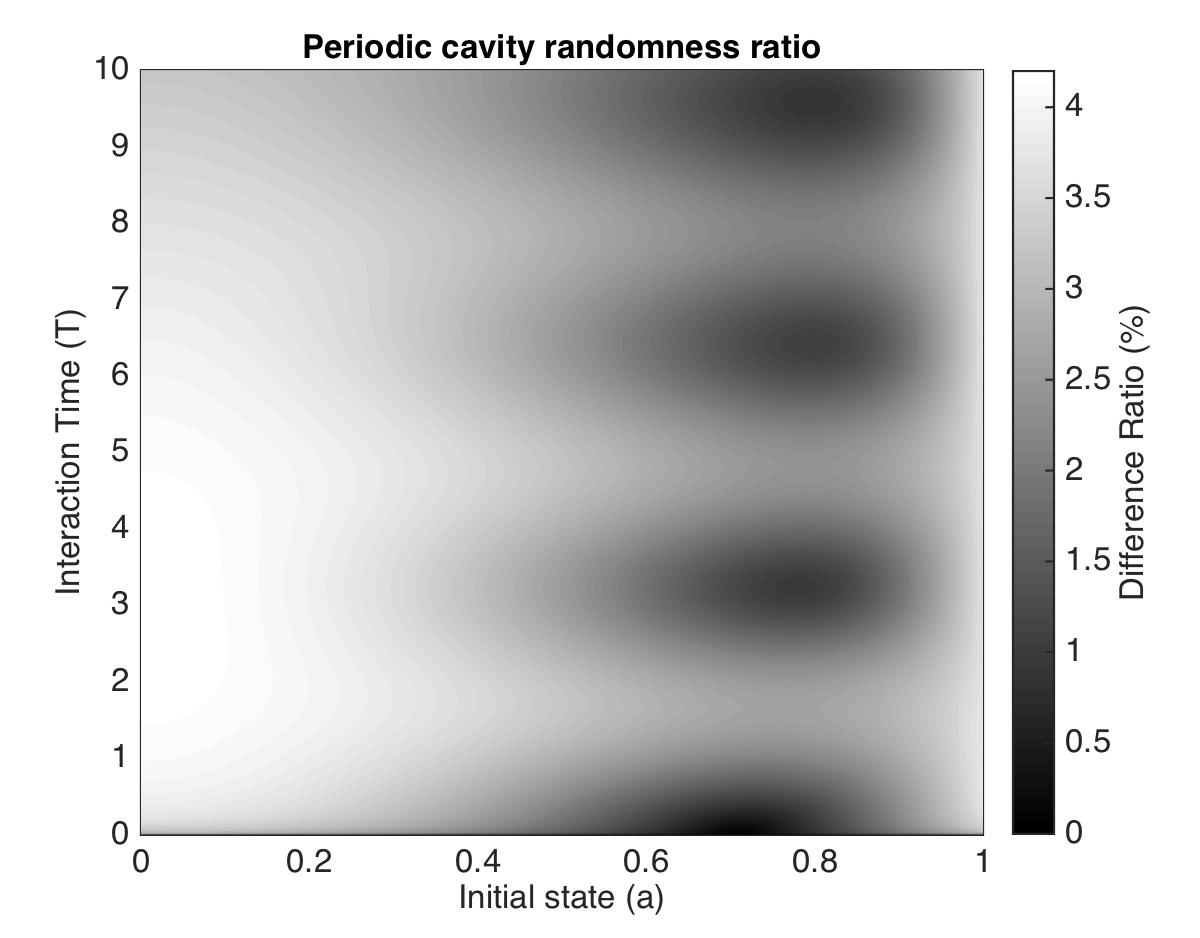}
\includegraphics[width=0.49\textwidth]{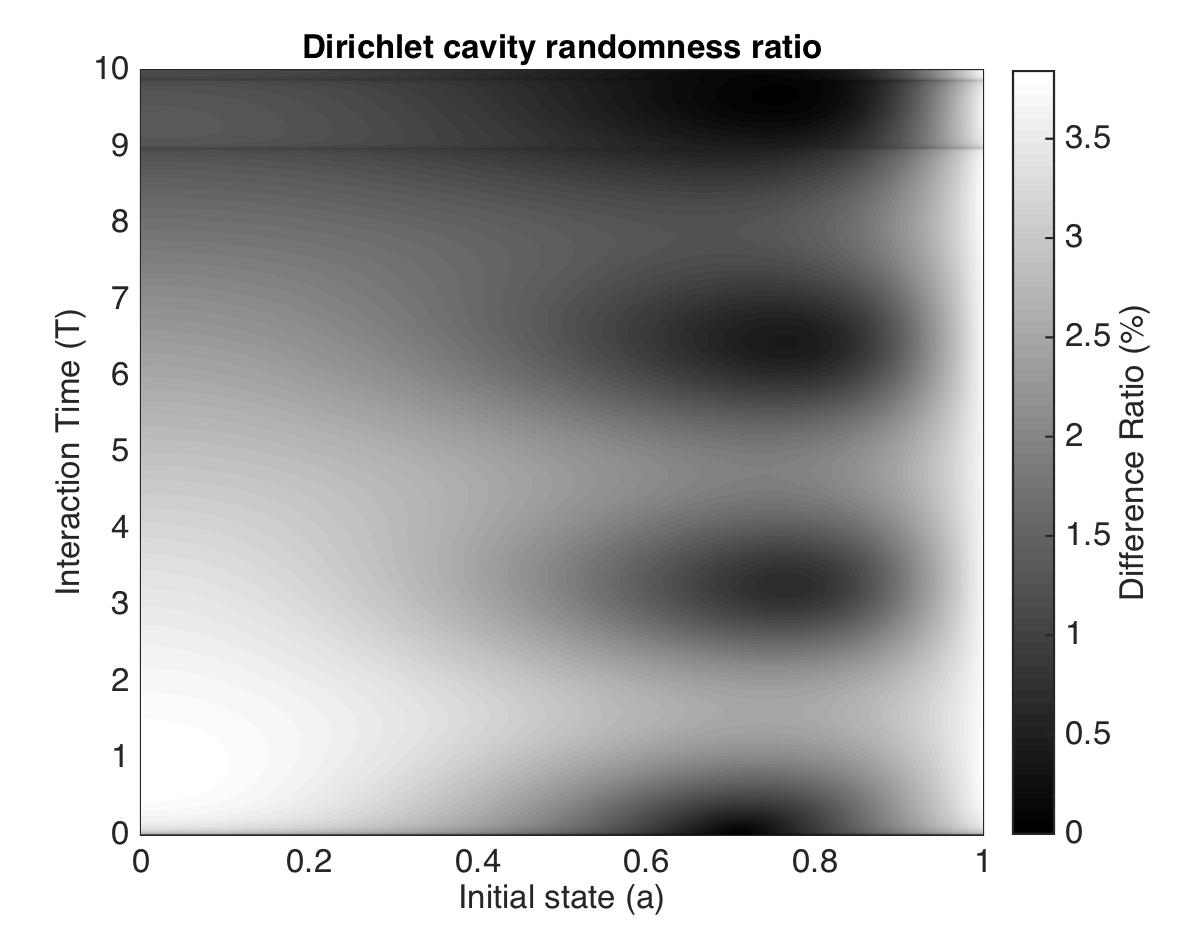}
\caption{\label{fig:ratios}Comparison between randomness computed for the full model and for the simplified RWA model in cavities, with chosen parameters $\lambda =0.01, \sigma=0.001, \Omega =1, m=3, N_c=6$, and $x_a=\pi L/6$. Notice the horizontal peaks in the figure on the right at the cavity light-crossing time due to the perturbation introduced by the switching returning to the original position of the atom after scattering with the boundaries. Note that $m=3$ corresponds to a Dirichlet cavity of length $L=3\pi$}
\end{figure*}

As seen in the figure~\ref{fig:ratios}, the randomness obtained from the fully relativistic calculation is \textit{lower} than the randomness computed from the rotating wave approximation model (i.e. $R\geq 0$). We interpret this as coming from the fact that non-relativisitic approximations (SMA and RWA) neglect all the correlations created between the atomic probe and the remaining of the field modes. Indeed, the shorter the interaction, the larger the bandwidth of the field modes that get perturbed by the interaction (this can be thought as a consequence of a time-energy uncertainty). 

%\begin{figure*}[t]
%\includegraphics[width=0.32\textwidth]{ratiovsT_excited_m3.jpg}
%\includegraphics[width=0.32\textwidth]{ratiovsT_plus_m3.jpg}
%\includegraphics[width=0.32\textwidth]{ratiovsT_ground_m3.jpg}
%\caption{\label{fig:ratiovsT}Randomness ratio as a function of interaction time for several initial states, with chosen parameters $\lambda=0.01, \sigma =0.001, \Omega=1, m=3, N_c=6, x_a=\pi L/6$.}
%\end{figure*}

\section{Conclusions\label{sec:conclusions}}

In this paper we considered an atom coupled for a short time with the electromagnetic field. By taking into account the full relativistic description of the atom-field interaction, we studied the amount of information shared as a result of this interaction by the atom and the field beyond the rotating-wave and single-mode approximations, as quantified by the guessing probability. We showed that small waiting times between preparation and measurement do not rid the atomic system from the problem of starting to share information with the quantized electromagnetic field to which the atom is unavoidably coupled. This is in stark contrast to what the usual approximated models of light-matter interaction predict.

In particular, the Jaynes-Cummings model under the single-mode approximation and the rotating-wave approximation would predict that the optimal way to proceed to reduce this entanglement --- and thus increase the randomness extractable from the atomic probe --- would be to prepare the ground state of the atom and the field. This is easy to understand already from a classical intuition:  If the atom is in the ground state and the field is not excited, the atom would remain in the ground state and thus it would not get correlated with the field. This intuition carries over to quantum optics under RWA and SMA. However, contrary to this intuition, we show that vacuum fluctuations entangle the atom with the field even in this case, and that this entanglement has significant consequences on the amount of certifiable randomness.

%By taking into account the full relativistic description of the atom-field interaction, our results show that it is generally not possible to extract as much as 1 bit of randomness out of an atomic qubit due to its unavoidable interaction with the electromagnetic field. Additionally, 

Also contrary to the classical intuition, we showed that the optimal amount of randomness is obtained for initial atomic states other than the ground state of the atom. This shows that the employment of the RWA and SMA in quantum optics does not provide a reliable lower bound on the amount of randomness that one can extract from an atomic probe.

As illustrative examples, we have analyzed the randomness loss due to these effects for the typical timescales and coupling regimes of strong and ultra-strong coupling in quantum optics and superconducting qubits in transmission lines, showing that the relative misestimation of the RWA and SMA models can indeed have a non-negligible magnitude.

Finally, our analysis suggests that the guessing probability as a function of the time of interaction converges to a constant value in some circumstances. If this result also carries to non-pertubative regimes this could allow for randomness certification independently of the interaction time. This is an interesting open question in its own right, and it will be studied elsewhere.

\appendix
\section{Relation between the SMA and RWA}
In this appendix, we discuss that the terms neglected by the SMA and RWA are of the same order. Therefore, both approximations are related.

Recall that we have a two-level atom interacting, according to Eq.~\eqref{eq:int}, with a quantized quantum field in periodic or Dirichlet cavity with the form of $\phi(x,t)$ given by Eq.~\eqref{eq:fieldfreespace} with the appropriate replacement rules. In this case, the dynamics is completely governed by the unitary time evolution operator which in the small coupling ($\lambda\ll \Omega$) is determined by Eq.~\eqref{afterfour}. Let us assume further that the atom is on-resonant with the cavity mode $\Omega=\omega_m$ for some fixed $m$.

%interaction Hamiltonian has the following form
%begin{align}
%    H_I^\text{SMA} \approx (a\sigma^+ + a^\dagger\sigma^-) + (a\sigma_- + a^\dagger\sigma^+).
%\end{align}
For illustration, the first order perturbation $U^{(1)}$ depends on terms of the form
\begin{equation}
\int_{-\infty}^{\infty}\d t \chi(t)\, e^{\pm i(\Omega\pm\omega_n)t},
\end{equation}
where the terms with $e^{\pm i(\Omega-\omega_n)t}$, corresponding to interaction terms of the form $a_n\sigma_+$ and $a_n^\dagger\sigma_-$, are usually called the rotating contributions, and the other terms where $e^{\pm i(\Omega+\omega_n)t}$ are called the counter-rotating contributions. Let us consider a constant interaction strength in some time interval $[t_{\text{start}},t_{\text{stop}}]$. Namely $\chi(t)=1$ in some $\Delta T=t_{\text{start}}-t_{\text{stop}}$ and $0$ elsewhere. The contribution of the resonant mode $m$ to the qubit's dynamics grows with $\Delta T$ while the off-resonant modes $n\neq m$ contribution stays bounded $\sim(\Omega-\omega_n)^{-1}$. The counter-rotating mode contributions are also always bounded $\sim(\Omega+\omega_n)^{-1}$.

%The second order perturbation $U^{(2)}$ depends on
%\begin{equation}
%\int_{-\infty}^{\infty}\d t \int_{-\infty}^{t}\d t' \chi(t)\chi(t') e^{\pm %i(\Omega\pm\omega_n)t} e^{\pm i(\Omega\pm\omega_n)t'},
%\end{equation}
%and under the same assumptions on $\chi(t)$ reduces to
%\begin{align}
%&\int_{t_{\text{start}}}^{t_{\text{stop}}}\d t \int_{t_{\text{start}}}^{t}\d t' e^{\pm i(\Omega\pm\omega_n)t} e^{\pm i(\Omega\pm\omega_n)t'}\\
%&= \int_{t_{\text{start}}}^{t_{\text{stop}}}\d t  e^{\pm i(\Omega\pm\omega_n)t} \frac{e^{\pm i(\Omega\pm\omega_n)t}-e^{\pm i(\Omega\pm\omega_n)t_{\text{start}}}}{\pm i(\Omega\pm\omega_n)}
%\end{align}
%messy calculations... I am not sure if this is the right path. Can we stop at first order and if so when? (perhaps $\lambda$ small enough?)

Thus if one make the assumption of SMA, namely dropping all contributions from off-resonant modes (because their contributions stay bounded while that of the resonant mode grows with interaction time $\Delta T$), then for consistency one must drop the contributions from the counter rotating terms since they are smaller, i.e. doing RWA as well.

For both approximations to be faithful already at leading order in perturbation theory we need to demand that
\begin{itemize}
    \item There is a resonant mode
    \item The interaction times are much larger than $\Omega^{-1}$.
\end{itemize}
Since the requirement $T\gg \Omega^{-1}$ is the same for both approximations, it is not a consistent approach (in these simple light-matter interaction models) to consider one and not the other without any further hypotheses. %These considerations at the level of the unitary evolution operator can be translated back to the level of the interaction Hamiltonian so that the fully relativistic model reduces to $\sim a\sigma_++ a^\dagger\sigma_-$.

%On the other hand, if one make the RWA, i.e. dropping terms with fast rotating exponent $\Omega+\omega_n$ then the model can still be multimode, namely $\sim \sum_n a_n\sigma_++ a_n^\dagger\sigma_-$.

\section*{Acknowledgments}

We thank Valerio Scarani, Jedrzej Kaniewski and Nicolas Sangouard for helpful discussions and correspondence. E. M-M. gratefully acknowledges Valerio Scarani for his hospitality during his visit to NUS-CQT.

This work is funded by the Singapore Ministry of Education (partly through the Academic Research Fund Tier 3 MOE2012-T3-1-009) and by the National Research Foundation of Singapore.

\bibliography{reference}

\end{document}